\documentclass[12pt]{article}
\usepackage{latexsym}
\usepackage{epsfig,amssymb,euscript,slashed}
\usepackage{amsmath}
\usepackage{cite} 
\usepackage{array,calc,epsfig}
\usepackage[linktocpage]{hyperref}
\usepackage{bbm}
\usepackage{fancybox}

\def\be{\begin{equation}}
\def\ee{\end{equation}}
\def\bseq{\begin{subequations}}
\def\eseq{\end{subequations}}

\def\bea{\begin{eqnarray}}
\def\eea{\end{eqnarray}}
\newcommand{\bbone}{\ensuremath{\mathbbm{1}}}
\newcommand{\ul}{\underline}
\def\bseq{\begin{subequations}}
\def\eseq{\end{subequations}}

\numberwithin{equation}{section} 
\usepackage[]{graphicx}

\def\d {{\rm d}}

\def\ie {{\em i.e.\ }}

\def\cala         {{\cal A}}

\def\calc         {{\cal C}}
\def\cald         {{\cal D}}

\def\calf         {{\cal F}}

\def\calh         {{\cal H}}

\def\calk         {{\cal K}}
\def\call         {{\cal L}}
\def\calm         {{\cal M}}
\def\caln         {{\cal N}}
\def\calo         {{\cal O}}
\def\calp         {{\cal P}}

\def\calr         {{\cal R}}
\def\cals         {{\cal S}}
\def\calt         {{\cal T}}
\def\calu         {{\cal U}}

\def\del          {\partial}
\def\delbar       {\bar\partial}
\def\ii           {{\rm i}}

\def\tr           {\mathop{\rm Tr}}
\def\Re           {{\rm Re\hskip0.1em}}
\def\Im           {{\rm Im\hskip0.1em}}

\def\sqr#1#2{{\vcenter{\vbox{\hrule height.#2pt
 \hbox{\vrule width.#2pt height#1pt \kern#1pt \vrule width.#2pt}\hrule
 height.#2pt}}}}

\def\d{\text{d}}

\def\slashchar#1{\setbox0=\hbox{$#1$}           
\dimen0=\wd0                                 
\setbox1=\hbox{/} \dimen1=\wd1               
\ifdim\dimen0>\dimen1                        
\rlap{\hbox to \dimen0{\hfil/\hfil}}      
#1                                        
\else                                        
\rlap{\hbox to \dimen1{\hfil$#1$\hfil}}   
/                                         
\fi}

\renewcommand{\bar}{\overline} 

\begin{document}
\font\cmss=cmss10 \font\cmsss=cmss10 at 7pt

\vskip -0.5cm
\begin{flushright}{\scriptsize ROM2F/2012/13 \\  \scriptsize  DFPD-12-TH-22}
\end{flushright}

\vskip .7 cm

\hfill
\vspace{18pt}
\begin{center}
{\LARGE \textbf{Brane instantons and fluxes in F-theory}}
\end{center}

\vspace{6pt}
\begin{center}
{\large\textsl{Massimo Bianchi$\,^{1}$, Gianluca Inverso$\,^1$ \& Luca Martucci$\,^{2,1}$}}

\vspace{25pt}
\textit{\small ${}^1$ Dipartimento di Fisica, Universit\`a di Roma ``TorVergata",\\ \& I.N.F.N. Sezione di Roma ``TorVergata'' ,  \\
Via della Ricerca Scientifica, 00133 Roma, Italy }\\  \vspace{12pt}

\textit{\small ${}^2$ Dipartimento di Fisica ed Astronomia ``Galileo Galilei",  Universit\`a di Padova,\\
\& I.N.F.N. Sezione di Padova,\\
Via Marzolo 8, 35131 Padova, Italy}
\end{center}

\vspace{12pt}

\begin{center}
\textbf{Abstract}
\end{center}

\vspace{4pt} {\small

\noindent We study the combined effect of world-volume and background fluxes on Euclidean D3-brane instantons in F-theory compactifications. We derive an appropriate form of the  fermionic effective action, in which the fermions are topologically twisted and the dynamical effect of fluxes, non-trivial axio-dilaton and warping is taken into account. We study  the structure of fermionic zero modes, which determines the form of possible non-perturbative superpotential and F-terms in the four-dimensional effective action.  Invariance under ${\rm SL}(2,\mathbb{Z})$ is discussed in detail, which allows for an interpretation of the results in terms of the dual M5-brane instanton in the M-theory picture.  We also provide the perturbative IIB description in the orientifold limit, when available.  
Furthermore, we consider the possible inclusion of  supersymmetry breaking bulk fluxes  and discuss its implications.
 }

\vspace{1cm}


\thispagestyle{empty}

\vfill
\vskip 5.mm
\hrule width 5.cm
\vskip 2.mm
{
\noindent  {\scriptsize e-mails:  {\tt massimo.bianchi@roma2.infn.it, gianluca.inverso@roma2.infn.it, luca.martucci@pd.infn.it} }
}

\newpage

\setcounter{footnote}{0}

\tableofcontents


\section{Introduction}

F-theory vacua \cite{Vafa:1996xn} are type IIB vacua characterized by the presence of  mutually non-local 7-branes, entailing a holomorphic axio-dilaton $\tau$ with non-trivial monodromy  along a  K\"ahler space $X$. A useful dual description of these vacua is provided by M-theory compactifications on elliptically fibered Calabi--Yau  $Y$, where $X$ and $\tau$ appear as the base of the fibration and the elliptic fiber complex structure respectively. In some cases, the F-theory vacua admit a limit in which they reduce to weakly coupled type IIB background with D7-branes and  orientifold $\Omega$7-planes \cite{sen1,sen2}.

 F-theory compactifications to four flat dimensions are obtained by taking $X$ six-dimensional and $Y$ eight-dimensional. They  represent a promising class of string compactifications, for reviews see e.g.\ \cite{Denef:2008wq,Weigand:2010wm}. In particular, they provide the starting point of phenomenologically appealing scenarios \cite{Kachru:2003aw,Balasubramanian:2005zx} in string theory. These settings crucially use two ingredients: the presence of {\em fluxes} which can be added in the internal space $X$ of the underlying F-theory background \cite{Becker:1996gj,GKP,Becker:2001pm} and provide a natural mechanism for stabilizing the complex structure, axio-dilaton and seven-brane moduli, for breaking supersymmetry at tree-level and for generating a non-trivial warping; the contribution of {\em non-perturbative effects} to the low-energy effective theory, in particular to the superpotential, which helps in stabilising the remaining moduli as well as in producing perturbatively forbidden couplings.   
 
 In particular, an important source of  non-perturbative corrections to the effective four-dimensional theory is provided by Euclidean D3-brane instantons, E3-brane instantons for short, which correspond to Euclidean M5-brane instantons in the dual M-theory picture.  Their effect on the low-energy effective theory in the absence of fluxes has been first studied in \cite{Witten:1996bn}. On the other hand, as recalled above, background fluxes play an important role in many interesting scenarios and it is crucial to understand how the results of \cite{Witten:1996bn} are affected by them. Furthermore, the generic E3-brane instanton can support itself a world-volume flux $\calf$ (dual to the M5-brane self-dual three-form flux). Since the effective low-energy superpotential, and hence vacuum structure and the low-energy effective theory, is ultimately determined by the sum of {\em all} relevant non-perturbative effects, the non-trivial world-volume must be incorporated into the analysis. Indeed, already in  the absence of bulk fluxes, a non-trivial world-volume flux on the E3-brane can significantly change the structure of the world-volume fermionic sector \cite{Bianchi:2011qh} (see \cite{Bianchi:2012pn} for a short summary) and lift some zero-modes which would prevent a flux-less E3-brane to contribute to the superpotential.\footnote{Of course, non-trivial world-volume fluxes have also other important physical effects, not necessarily related to
to the counting of fermionic zero modes, as for instance the ones  discussed in \cite{Grimm:2011dj,Kerstan:2012cy}.} The presence of bulk fluxes makes the  incorporation of a non-trivial $\calf$ in the discussion  even more compulsory,  since the world-volume Bianchi identity relates its exterior derivative, $\d \calf $, to the bulk three-form flux. Hence, if the latter is non-vanishing, $\calf$ is generically non-vanishing as well. 
 
The nature of the low-energy effective terms produced by instantons depends on the dynamics of fermionic quantum fluctuations around the instantonic configuration. In particular, the fermionic zero-modes  determine which fermions must be inserted in the path-integral in order to produce a non-vanishing effective interaction. Hence, in order to understand the effect of E3-brane instantons (or the dual M5-branes) in F-theory flux compactifications it is important to have a suitable form for the effective theory governing the fermionic quantum fluctuations of the E3-brane  and to understand how the nature of the associated zero-modes is affected by the bulk and world-volume fluxes. This is the main goal of the present paper.     

We will focus on the neutral fermionic sector, associated with E3-E3 open strings. Hence, we will not discuss the possible role of charged string modes which can arise at the intersection of the E3-brane with background branes and which can be the source of a dependence of the non-perturbative effect on
the fields associated to the background branes, in particular on charged chiral matter -- for more details and more complete lists of references  see for instance the review papers \cite{Blumenhagen:2009qh,Bianchi:2009ij}.  Starting from the (Wick-rotated) Green-Schwarz formulation of the E3-brane effective action, we will work in the semiclassical one-loop approximation, considering the fermionic effective action at the quadratic level in the fermionic fluctuations but including the complete dependence on the bulk geometry and on the classical world-volume configuration. (Higher order terms in the fermions, which may play a role in lifting zero-modes of the quadratic action, are harder to determine.) We will mostly work directly in type IIB, partly because the E3-brane effective action is somewhat  simpler than  the dual M5-brane one, partly because in this way the results obtained in the supergravity description may be more easily compared and complemented by results obtained with string world-sheet techniques.  In any case, we eventually describe an uplift of our results to the M5-brane in the dual M-theory picture, which can be more useful for addressing some non-trivial topological aspects.

The effect of bulk and world-volume fluxes on the fermionic sector of E3-brane or M5-brane instantons in F-theory/IIB orientifold backgrounds has been previously studied in the literature. The effect of bulk fluxes on the counting of fermionic zero modes of \cite{Witten:1996bn} has been considered in \cite{Saulina:2005ve,Kallosh:2005gs,Bergshoeff:2005yp,Lust:2005cu,Tsimpis:2007sx} under the simplifying assumption of a vanishing world-volume flux (assumption which, as discussed above, is generically not admissible) and neglecting warping effects.  Other papers studying various related aspects  from different perspectives are \cite{Gorlich:2004qm,Tripathy:2005hv,Blumenhagen:2007bn,GarciaEtxebarria:2008pi,Uranga:2008nh,Billo:2008sp,Billo:2008pg}.\footnote{See also \cite{Blumenhagen:2010ja,Cvetic:2010rq,Donagi:2010pd,Marsano:2011nn,Cvetic:2011gp,Cvetic:2012ts} for other recent work on E3-brane instantons in F-theory compactifications.} One of the aims of the present paper is to present a systematic analysis of the problem, completing and unifying the partial results previously presented in the literature. This work may be considered as a natural follow-up and completion of \cite{Bianchi:2011qh}, which focused on `magnetized' (or `fluxed') E3-branes on flux-less backgrounds, partly using type IIB SL(2,$\mathbb{Z}$) duality  as guiding principle. As we will see, the systematic analysis presented in this paper highlights some interesting subtleties related to the SL(2,$\mathbb{Z}$) duality, which  are however negligible at the practical level for the most immediate physical applications.  

\bigskip

In order to guide the reader through  this lengthly and at times unavoidably technical paper, we now provide a schematic description of its structure. 

In Section \ref{sec:sugraback} we review the structure of F-theory flux backgrounds in their IIB formulation. It provides the background material to precisely interpret the results of the subsequent sections. In particular, we recall how the non-trivial non-perturbative structure F-theory backgrounds is encoded in non-trivial monodromies of the SL(2,$\mathbb{Z}$) duality group. 

In Section \ref{sec:E3bos}
we carefully spell out the supersymmetry conditions on the classical E3-brane instantonic configurations and their relevant properties.

 In Section \ref{sec:fermeff} we present one of the main results of the paper: the quadratic effective action for the world-volume fermions. In this action the fermions are naturally topologically twisted along the E3 world-volume and the complete effect of bulk and world-volume fluxes, axio-dilaton and warping, is explicitly exhibited.  
 
 Sections \ref{sec:actionduality}, \ref{sec:offshell} and \ref{sec:pert} concern the properties of the world-volume fermionic theory  under the type IIB SL(2,$\mathbb{Z}$) duality group and under the orientifold projections, which are essential for understanding the global properties of the action. In particular, we will discuss a subtlety on SL(2,$\mathbb{Z}$) duality,  which turns out to be related to the observation that the supersymmetric E3-brane configurations do not generically solve the embedding equations of motion because of the non-trivial bulk structure. 
 
 In Section \ref{sec:zm} and \ref{sec:geomint} we study in detail the structure of the fermionic zero modes, showing how some of them can be lifted by the combination of bulk and world-volume fluxes and providing a geometrical interpretation of part of these zero mode lifting effects.
 
  In Section \ref{sec:supgen} we outline some possible applications of the results presented in the previous sections to the computation of F-terms in the low-energy effective action, briefly reviewing and adapting some of the standard procedures followed in similar contexts.  

In Section \ref{sec:SB} we discuss in what sense and under which conditions  the world-volume fermionic terms associated with a possible supersymmetry breaking bulk flux agree with what expected from the low-energy effective theory point of view. 

In Section \ref{sec:uplift} we uplift our results to the M5-brane in the dual M-theory background. As recalled above, the M-theory perspective 
can be very useful for understanding relevant global topological properties, which are more difficult to describe in the type IIB framework.
In particular, we will see that the subtleties with the    SL(2,$\mathbb{Z}$) duality encountered with the E3-brane translate into the well known difficulty of writing an off-shell covariant effective action for the M5-brane (without introducing auxiliary fields \cite{Pasti:1997gx,Bandos:1997ui}). 

The appendices contain a summary of the conventions and some technical details, in particular a description of the derivation of the effective action presented in section \ref{sec:fermeff}. 


\section{F-theory compactifications and fluxes}
\label{sec:sugraback}

In this section we review the structure of F-theory vacua in the presence of fluxes, fixing our notation and conventions. 
This is well know material and the expert reader can skip it and jump directly to the next section.

F-theory vacua are type IIB non-perturbative backgrounds characterized by the presence of D7-branes and of their ${\rm SL}(2,\mathbb{Z})$ dual $(p,q)$ 7-branes. In the absence of fluxes, F-theory compactifications to four dimensions are dual to the uplift of M-theory compactifications on elliptically fibered Calabi-Yau (CY) four-folds \cite{Vafa:1996xn}. These preserve minimal $\caln=1$ supersymmetry in four-dimensions and admit both supersymmetry preserving \cite{Becker:1996gj} and supersymmetry breaking \cite{Becker:2001pm} fluxes.    

Although the M-theory description provides an elegant and efficient geometrization of these spaces, in this paper 
we work directly in type IIB, as in \cite{Bianchi:2011qh}. This perspective is  more directly related to the perturbative superstring description, which may then be used to complete the results obtained in the supergravity regime, at least 
when a weak coupling orientifold limit  exists \cite{sen1,sen2}.

\subsection{Bulk supergravity conditions}
\label{sec:bulksugra}

We now summarize the conditions defining the IIB backgrounds \cite{GKP} we are interested in.
In the Einstein frame, the metric can be written as
\be\label{Emetric}
\d s^2_{\rm E}=e^{2A}\d x^\mu\d x_\mu+e^{-2A}\d s^2_X
\ee
where $\d x^\mu\d x_\mu$ denotes the four-dimensional Minkowski metric, $\d s^2_X$ is a K\"ahler metric 
on the internal six-dimensional manifold and the warping $e^A$ in general depends on the internal (real) coordinates $y^m$. Being $X$ a K\"ahler space, we can introduce complex coordinates, denoted by $z^i$, $i=1,2,3$, together with their complex conjugates $\bar z^{\bar\imath}$, and K\"ahler form $J=-\ii g_{i\bar\jmath} \d z^i\wedge \d \bar z^{\bar\jmath}$.

The presence of 7-branes is signaled by a non-trivial axio-dilaton $\tau\equiv C_{\it 0}+\ii e^{-\phi}$,\footnote{In our notation, $e^\phi$ is the dilaton, $H_{\it 3}=\d B_{\it 2}$ is the NS-NS three-form, and the R-R fluxes can be packed into the polyform  $F=\sum^4_{{\it k}=1}F_{\it 2k+1}=(\d+H\wedge)C$, with $C=\sum^4_{{\it k}=0}C_{\it 2k}$ being the R-R potentials.} which must be holomorphic, namely
\be\label{holtau}
\delbar\tau=0
\ee  
where $\delbar=\d\bar z^{\bar\imath}\wedge\bar\del_{\bar\imath}$ is the standard Dolbeault operator (and $\del=\d z^{i}\wedge\del_{i}$ is its complex conjugate, so that $\d=\del+\delbar$).

One can introduce an imaginary-self-dual (ISD) three-form flux $G_{\mathit 3}=F_{\it 3}+\ii e^{-\phi}H_{\it 3}$ (this is often written as $G_{\it 3}=\tilde F_{\it 3}+\tau H_{\it 3}$, where $\tilde F_{\it 3}=F_{\it 3}-C_{\it 0}H_{\it 3}=\d C_{\it 2}$):
\be
*_X G_{\it 3}=\ii G_{\it 3}
\ee
By using the K\"ahler structure on $X$, an ISD flux can be decomposed  as $G_{\it 3}=G_{2,1}+G_{0,3}$, with $G_{0,3}=\frac{1}{3!}G_{\bar\imath\bar\jmath\bar k}\d \bar z^{\bar\imath}\wedge \d\bar z^{\bar\jmath}\wedge\d \bar z^{\bar k}$ and $G_{2,1}=\frac{1}{2}G_{ij\bar k}\d z^i\wedge \d z^j\wedge \d\bar z^{\bar k}$, where $G_{2,1}$ is also primitive, namely $J\wedge G_{2,1}=0$. The primitive $(2,1)$ component preserves the background (four-dimensional $\caln =1$) supersymmetry, while the $(0,3)$ component breaks supersymmetry but still solves the classical supergravity equations.%
\footnote{An IAS $G_{\it 3}$ could also  have a non-primitive (1,2) component, which is however absent in the F-theory compactifications considered in this paper.} 

Finally, there is also a self-dual five-form flux $F_{\it 5} = \d C_{\it 4} + H_{\it 3}\wedge C_{\it 2}$, which is related to the warping by the following formula
\be\label{F5}
F_{\it 5} = \d x^{0123}\wedge \d e^{4A}  +   *_X\d e^{-4A}
\ee
where $*_X$ uses the K\"ahler metric $\d s^2_X$.
The associated Bianchi identity gives a Poisson-like equation for the warping
\be\label{warping}
\Delta_X e^{-4A}=\frac{1}{2\Im\tau} G_{\it 3}\cdot \bar G_{\it 3} +\delta_{\rm D3}
\ee
where $\Delta_X$ is the Laplacian computed using the K\"ahler metric on $X$, $G_{\it 3}\cdot \bar G_{\it 3}=\frac{1}{3!}G_{mnp}\bar G^{mnp}$ and  $\delta_{\rm D3}$ represents the localized D3-charge, either on space-filling D3-branes, or $\Omega$3-planes, or induced on 7-branes.  This equation can be always integrated and uniquely determines $e^{-4A}$ up to an additive constant $c$:
\be\label{breathing}
e^{-4A}=c+e^{-4\hat{A}}
\ee
where $e^{-4\hat{A}}$ is a given solution of the equation (\ref{warping}). The constant $c$ represents the universal breathing  mode of this class of compactifications. On the other hand, an overall rescaling of the internal metric $\d s^2_X$ is not a modulus, unless it is reabsorbed by a rescaling of the warping which preserves $e^{-2A}\d s^2_X$. Such a rescaling can be seen as a (global) Weyl rescaling of the four-dimensional Poincar\'e metric and should not be considered as a modulus \cite{Giddings:2005ff}. Hence, we can fix the overall normalization of the internal metric $\d s^2_X$ by imposing, for instance, that it measures a volume of $X$ of order one. Here and in the rest of the paper we use natural units such that  $2\pi\sqrt{\alpha'}=1$. 

\subsection{${\rm SL}(2,\mathbb{Z})$ duality}
\label{sec:bulkduality}

By definition, F-theory backgrounds are characterized by a holomorphic axio-dilaton $\tau$ with non-trivial monodromy. Indeed, F-theory vacua admit a perturbative IIB description only locally. Globally, the internal space should be thought of as covered by different patches, glued together via non-trivial ${\rm SL}(2,\mathbb{Z})$ duality transformations, under which
\be\label{tauduality}
\tau\quad\rightarrow\quad \frac{a\tau+b}{c\tau+d}
\ee 
with $a,b,c,d\in\mathbb{Z}$ and $ad-cb=1$. 

The {\em Einstein frame} metric and the R-R 5-form $F_{\it 5}$ are inert under the duality (\ref{tauduality}). 
On the other hand, $\tilde F_{\it 3}\equiv F_{\it 3}-C_{\it 0}H_{\it 3}$ and $H_{\it 3}$ transform as a doublet: 
\be\label{3formduality}
\left(\begin{array}{c} \tilde F_{\it 3} \\ H_{\it 3} \end{array}\right)\quad \rightarrow \quad \left(\begin{array}{cc} a & -b \\ -c & d \end{array}\right)
\left(\begin{array}{c} \tilde F_{\it 3} \\ H_{\it 3} \end{array}\right)
\ee
This transformation rule can be conveniently re-expressed in terms of the complex combination $e^{\frac{\phi}{2}}G_{\it 3}$, which transforms by a phase shift
\be
e^{\frac{\phi}{2}}G_{\it 3}\quad\rightarrow\quad e^{-\ii \alpha}\,e^{\frac{\phi}{2}}G_{\it 3}
\ee  
where $\alpha=\arg(c\tau+d)$. 

As we will see, with the exception of $\tau$, the transformation properties of all the other fields we will encounter  can be described by a similar phase shift. Such transformations can be seen as U(1) gauge transformations and one can actually use the axio-dilaton $\tau$ to construct an abelian connection 
\be\label{Qconn}
Q\equiv Q_m\d y^m=\frac{1}{2}e^\phi F_{\it 1}=-\frac{\ii}{2}(\del-\delbar)\phi
\ee
Hence we denote the associated gauge group by U$(1)_Q$.  In (\ref{Qconn}) $F_{\it 1}=\d C_{\it 0}$ and the last equality holds when $\bar\partial\tau = 0$.

More precisely, we say that a field $\Phi$ has U$(1)_Q$ charge $q_Q$ if it transforms as
\be
\Phi\qquad\rightarrow \quad e^{\ii q_Q\alpha}\Phi\quad~~~~~ \text{with}\quad \alpha=\arg(c\tau+d)
\ee 
under ${\rm SL}(2,\mathbb{Z})$ duality. Then the associated covariant derivative is 
\be
\nabla^Q_m\Phi\equiv(\nabla_m-\ii q_Q Q_m)\Phi
\ee
Hence, for instance, $e^{\frac{\phi}{2}}G_{\it 3}$ has $q_Q=-1$. One can also check that $\del\phi=\frac{\ii}{2}e^\phi\d\tau$ has $q_Q=-2$ and 
that the Bianchi identities for $F_{\it 3}$ and $H_{\it 3}$ can be compactly written as  $\d_Q(e^{\frac{\phi}{2}}G_{\it 3})=e^{\frac{\phi}{2}}\del\phi\wedge \bar G_{\it 3}$, which is manifestly ${\rm SL}(2,\mathbb{Z})$ invariant. 

 Since $(\d Q)_{0,2}=0$,  $Q$ actually defines a holomorphic line bundle $\call_Q$ which is isomorphic to the the inverse of the holomorphic canonical bundle $\calk^{-1}_X$ since $\call_Q\otimes \calk_Q$ is a trivial line bundle admitting a global section.\footnote{See for instance section 2 of \cite{Bianchi:2011qh} for more details on the type IIB description of F-theory backgrounds and its relation with the dual M-theory picture.} Charge $q_Q$ fields can be thought of as sections of $\call_Q^{q_Q}$.

\subsection{Supersymmetric structure}

We already mentioned that a four-dimensional $\caln=1$ supersymmetry is preserved if $G_{\it 3}$ has only primitive (2,1) component. 
We now summarize some useful facts about the preserved supersymmetric structure. For later purposes, it is convenient to adopt the {\em string} frame, in which the ten-dimensional metric is $\d s^2=e^{\frac{\phi}{2}}\d s^2_{\rm E}$, with $\d s^2_{\rm E}$ given in (\ref{Emetric}).

First, let us split the ten-dimensional gamma matrices $\Gamma^{\ul M}$ according to the split of space-time into four-plus-six dimensions%
\footnote{Underlined indices refer to the locally flat frame defined by the vielbein.}
\be
\begin{aligned}\label{def Gamma}
\Gamma^{\ul \mu}&=\tilde\gamma^{\ul \mu}\otimes \bbone\quad~~~~(\ul\mu=0,\ldots,3)\\
\Gamma^{{\ul{3+ m}}}&=\tilde\gamma_5\otimes \gamma^{\ul m}\quad~~~(\ul m=1,\ldots,6)
\end{aligned}
\ee
Here $\tilde\gamma_5=\ii\tilde\gamma^{\ul{0123}}$ is the four-dimensional chirality operator, while $\gamma_7=-\ii\gamma^{\ul{1\ldots6}}$ and $\Gamma_{11}=\Gamma^{\ul{0\ldots 9}}$ are the six-dimensional and ten-dimensional ones respectively, so that $\Gamma_{11}=\tilde\gamma_5\otimes \gamma_7$. 
We use the Majorana-Weyl (MW) representation in which the gamma matrices  $\Gamma^{\ul M}$ and $\tilde\gamma^{\ul\mu}$ are real while $\gamma^{\ul m}$ are purely imaginary.  The type IIB supersymmetry generators are given by two  spinors $\epsilon_1$ and $\epsilon_2$ which are Majorana-Weyl (MW), namely they are real and such that $\Gamma_{11}\epsilon_{1,2}=\epsilon_{1,2}$.
When $G_{0,3}=0$, the background admits four independent  Killings spinors, which have the form $ \epsilon_1+\ii\epsilon_2=\epsilon_{\rm R}$ and $ \epsilon_1-\ii\epsilon_2=\epsilon_{\rm L}$, with
\begin{equation}\label{killing}
 \epsilon_{\rm R}= e^{\frac{A}{2}+\frac{\phi}{8}}\,\zeta_{\rm R} \otimes \eta\qquad~~~~~~~~\epsilon_{\rm L}=e^{\frac{A}{2}+\frac{\phi}{8}}\,\zeta_{\rm L}\otimes \eta^* 
\end{equation}
where $\zeta_{\rm R}$ is a right-handed ($\tilde\gamma_5\zeta_{\rm R}=\zeta_{\rm R}$)  constant spinor in four-dimensions, $\zeta_{\rm L}=\zeta^*_{\rm R}$ is its complex conjugate of opposite chirality, and $\eta$ is chiral  in six dimensions ($\gamma_7\eta=\eta$). The spinor $\eta$ has U$(1)_Q$ charge $q_Q=\frac12$ and satisfies the U(1)$_Q$-covariant Killing equations
\begin{equation}\label{kileq}
\nabla_m^Q\eta\equiv (\nabla_m  -\frac\ii 2 Q_m )\eta =0.
\end{equation}
Furthermore, we normalize   $\eta^\dagger\eta=1$. 

With the internal spinor $\eta$ one can construct the following forms which characterize the internal K\"ahler space:
\begin{equation}\label{SU(3)def}
 J = \frac{\ii}{2} \eta^\dagger \gamma_{mn} \eta\,\d y^m\wedge\d y^n \qquad~~~
 \Omega =\frac{1}{3!}e^{\frac{\phi}{2}} \eta^T \gamma_{mnp} \eta\, \d y^m\wedge\d y^n\wedge\d y^p
\end{equation}
where the curved gamma matrices are those associated with the metric $\d s^2_X$ in (\ref{Emetric}).
$J$ is just the K\"ahler form already introduced after (\ref{Emetric}), while  $\Omega$ is a holomorphic $(3,0)$ form: {\em i.e.}\ $\delbar\Omega=0$. $J$ and $\Omega$ are related by the following normalization condition
\be
\frac{1}{3!}\, J\wedge J\wedge J=-\frac{\ii}{8}\, e^{-\phi}\Omega\wedge \bar\Omega
\ee
Notice that $\Omega$ is not covariantly constant. This can be understood by observing that  $e^{-\frac{\phi}{2}}\Omega$ has U$(1)_Q$ charge $q_Q=1$ and indeed it satisfies the appropriate U$(1)_Q$ covariant condition   
\be\label{omegaeq}
\nabla^Q_m(e^{-\frac{\phi}{2}}\Omega)=0
\ee

Restating these results in terms of holomorphic line-bundles, $\Omega$ is a holomorphic global section of the holomorphic line bundle $\calk_X\otimes\call_Q$, which is trivial as recalled at the end of section \ref{sec:bulkduality}.   Clearly, in the case of constant axio-dilaton, the internal space reduces to a standard CY (with trivial $\calk_X$).

\section{Introducing E3-branes instantons}
\label{sec:E3bos}

Let us now turn our attention to our  main subject, namely Euclidean D3-branes, or E3-branes for short.  
One is interested in these objects because they can generate perturbatively forbidden terms in the four-dimensional low-energy effective theory. 
In particular, supersymmetric configurations can contribute to the four-dimensional superpotential. We will expand on this point in section \ref{sec:supgen}. In this section we mainly describe general properties of the E3-branes.

 We will focus on E3-branes which wrap a four-dimensional submanifold of the internal space $X$ and sit at a certain
point $x_0^\mu$ of the external flat space. Hence, from the four-dimensional point of view, such configurations appear as point-like and then they are also referred to as E3-brane instantons. In particular, we are interested in E3-branes which preserve some supersymmetry. Hence, in the following we will mostly work on supersymmetric backgrounds, that is with $G_{0,3}=0$, explicitly mentioning when we include a non-vanishing $G_{0,3}$ in the discussion. 

The E3-brane supersymmetry conditions are obtained by $\kappa$-symmetry arguments, see for instance \cite{Bianchi:2011qh} or the more general discussion of appendix D of \cite{eff1}. Here we briefly summarize the main steps which can be useful in the following sections.

Let us first remark that, as usual in instanton calculations, we are actually working in the Wick rotated version of the supergravity background described in section \ref{sec:sugraback}, in which the Minkowski four-dimensional space-time becomes the Euclidean $\mathbb{R}^{4}$. The key point is that the Wick rotation acts on the Killing spinor  (\ref{killing}) by analytically continuing  $\zeta_{\rm R}$ and $\zeta_{\rm L}$ to two independent spinors which are not related by complex conjugation but should each still be considered as counting two independent supersymmetries.  As a result E3-brane instantons can preserve either $\epsilon_{\rm R}$ or $\epsilon_{\rm L}$. We focus on the first case while the latter simply corresponds to  an {\em anti}-E3-brane instanton and produces effects which are `complex conjugate' to the ones produced by an E3-brane instanton. 

The supersymmetry conditions are encoded in the spinorial projection condition
 \be\label{gammasusy}
 \Gamma_{\rm E3}\,\epsilon_{\rm R}=\ii\epsilon_{\rm R}
 \ee
where 
\be\label{proj}
\Gamma_{\rm E3}=\frac{\ii\, \epsilon^{A_1\ldots A_4}}{\sqrt{\det(g|_D+\calf)}}\Big(\frac{1}{4!}\Gamma_{A_1\ldots A_4}+\frac{1}{4}\Gamma_{A_1A_2}\calf_{A_3A_4}+\frac{1}{8}\calf_{A_1A_2}\calf_{A_3A_4}\Big)
\ee
Let us explain our notation. The world-volume of the E3-brane is parametrized by coordinates $\sigma^A$, $A=1,\dots,4$ and $\Gamma_{A_1\ldots A_p}$ are the pullback of the antisymmetrized ten-dimensional gamma matrices  $\Gamma_{M_1\ldots M_p}\equiv\Gamma_{[M_1}\cdots\Gamma_{M_p]}$ to the cycle $D$ which is wrapped by the E3-brane. Otherwise, in index free notation, we denote the pull-back to $D$ by $|_D$. Hence,
$g|_D$ is the pull-back of the bulk string-frame metric. Finally, $\calf$ is the gauge-invariant world-volume flux. In the standard formulation of supersymmetric D-brane effective actions \cite{Cederwall:1996ri,Bergshoeff:1996tu}, which we are Wick rotating for our purposes, the world-volume flux $\calf$ must satisfy the Bianchi identity $\d\calf=H_{\it 3}|_D$, which means that we can locally write $\calf=\frac{F_{\rm wv}}{2\pi}+B_{\it 2}|_D$ for some world-volume U(1) field-strength $F_{\rm wv}=\d A_{\rm wv}$. Hence, in the presence of bulk fluxes,  $\calf$ is generically non-trivial and there is no freedom to turn it off.  

One can then show  that (\ref{gammasusy})  boils down to two conditions on the E3-brane configuration. The first requires the E3-brane to wrap an holomorphically embedded four-cycle, alias {\em divisor},  $D\subset X$. The second requires  the world-volume two-form flux $\calf$ to be self-dual:
\be\label{SDflux}
\calf=*_D\calf\quad~~~~~\Leftrightarrow\quad~~~~~\calf_{2,0}= \calf_{0,2}=0\quad\text{and}\quad J|_D\wedge \calf=0
\ee
where $J|_D$ is the K\"ahler form on $D$ induced by the bulk.
In other words, $\calf$ is purely $(1,1)$ and primitive.

Since the bulk space is K\"ahler, the divisor $D$ is actually calibrated, in the sense that $\sqrt{\det h}\, \d^4\sigma=-\frac12(J\wedge J)|_D$,
where $h$ denotes the pull-back of the bulk K\"ahler metric $\d s^2_X$ to $D$, that is $g|_D\equiv e^{-2A+\frac{\phi}{2}}h$. 
Actually, the combination of the holomorphy of the embedding and of (\ref{SDflux}) is equivalent to a generalized calibration in sense of \cite{Koerber:2005qi,luca1}:
\be
e^{-\phi}\sqrt{\det(g|_D+\calf)}\, \d^4\sigma=-\frac{1}{2}e^{-4A}(J\wedge J)|_D+\frac{1}{2}e^{-\phi}\calf\wedge \calf
\ee
This implies that the E3-brane effective action 
\be\label{E3action1}
S_{\rm E3}=2\pi\int e^{-\phi}\sqrt{\det(g|_D+\calf)}-2\pi\ii \int C\wedge e^\calf
\ee
 simplifies drastically once evaluated on  supersymmetric configurations:
\be\label{E3action2}
S^{\rm susy}_{\rm E3}=2\pi\int_D\Big[-\frac{1}{2}e^{-4A}J\wedge J-\frac{\ii}{2}\,\tau\calf\wedge\calf-\ii C_{\it 2}\wedge \calf-\ii C_{\it 4}\Big]
\ee

The supersymmetric conditions guarantee that the field-strength $\calf$ transforms well under $\text{SL}(2,\mathbb{Z})$ duality transformations. 
In general, $\text{SL}(2,\mathbb{Z})$ acts on the world-volume flux as an electromagnetic  duality \cite{Tseytlin:1996it,Green:1996qg}. Namely, one can define a dual $\calf_{\rm D}$ world-volume flux by varying $\calf$ in the action and then setting $\delta S_{\rm E3}=\ii\int (\calf_{\rm D}-C_{\it 2})\wedge \delta\calf$.
Then, $(\calf_{\rm D},\calf)$ transform as a doublet under $\text{SL}(2,\mathbb{Z})$, exactly as the pair $(\tilde F_{\it 3},H)$ in (\ref{3formduality}),
and one can construct the following complex combinations with definite U(1)$_Q$ charge:
\be\label{sdfluxes}
\begin{aligned}
e^{-\frac{\phi}{2}}\calf_+&\equiv\frac{\ii}{2}e^{\frac{\phi}{2}}(\calf_{\rm D}+\bar\tau \calf) \qquad\qquad\quad (q_Q=+1)\\
e^{-\frac{\phi}{2}}\calf_-&\equiv-\frac{\ii}{2}e^{\frac{\phi}{2}}(\calf_{\rm D}+\tau \calf) \qquad\qquad (q_Q=-1)
\end{aligned}
\ee
If we compute $\calf_{\rm D}$ for a supersymmetric E3-brane, one simply finds $\calf_{\rm D}=-\tau\calf$, so that $\calf_-=0$ and $\calf=\calf_+$.
Hence, one can conclude that $e^{-\frac{\phi}{2}}\calf$ transforms with U(1)$_Q$ charge $q_Q=+1$ under $\text{SL}(2,\mathbb{Z})$.

It may be useful to rewrite (\ref{E3action2}) in a more manifestly duality covariant form. First of all, recall that $C_{\it 4}$ transforms  under  $\text{SL}(2,\mathbb{Z})$
in such a way that $F_{\it 5}=\d C_{\it 4}+H_{\it 3}\wedge C_{\it 2}$ remains invariant. Then it is useful to introduce a modified R-R potential 
\be\label{invRR4}
\tilde C_{\it 4}\equiv C_{\it 4}+\frac12 B_{\it 2}\wedge C_{\it 2}
\ee
which is invariant under  $\text{SL}(2,\mathbb{Z})$, since $F_{\it 5}=\d \tilde C_{\it 4}+\frac 12 (H_{\it 3}\wedge C_2-\tilde F_3\wedge B_{\it 2})$. 
Hence, the action (\ref{E3action2}) can be rewritten as
\be\label{E3action3}
\begin{aligned}
S^{\rm susy}_{\rm E3}=&2\pi\int_D\Big(-\frac{1}{2}e^{-4A}J\wedge J-\ii \tilde C_4\Big)+\ii\pi\int_D\Big(\calf_{\rm D}\wedge B_{\it 2}-\calf\wedge C_{\it 2}\Big)+\frac{\ii}{4\pi}\int_D  F^{\rm wv}\wedge F^{\rm wv}_{\rm D}
\end{aligned}
\ee
where $F^{\rm wv}=\d A^{\rm wv}$ is the U(1) world-volume field strength, while $F^{\rm wv}_{\rm D}$ is its electromagnetic dual. 
 The first two terms on the right-hand side of (\ref{E3action3}) are manifestly invariant under $\text{SL}(2,\mathbb{Z})$ while the last term is 
 the appropriate non-invariant one required by the general arguments of \cite{Gaillard:1981rj}.\footnote{\label{foot:bosE3} The actions (\ref{E3action1}), (\ref{E3action2}) and (\ref{E3action3}) are not literally valid for E3-branes in F-theory vacua. Indeed, they  generically hold only on local patches of $D$ and a proper global definition must take into account the possible SL(2,$\mathbb{Z})$ duality transformations connecting different patches, for instance along the lines of \cite{Ganor:1996pe}. In this paper, we are not going to enter into the details of the proper global definition of the E3 bosonic action but keep in mind that (\ref{E3action1}) and (\ref{E3action2}) must be appropriately completed.}  

\section{E3-brane fermionic effective action}
\label{sec:fermeff}

As in  standard (supersymmetric) instanton calculations, the possible effects  of brane instantons on the low-energy effective theory
are substantially characterized by the fermionic sector of the theory and by the associated zero modes. In   
the one-loop approximation, zero-modes drop from  the quadratic action and must be soaked up by external insertions in the path-integral. In fact, higher order terms may help lifting zero-modes of the quadratic action buy they are harder to find. In order to obtain the action which governs the fermionic E3-brane fluctuations around the supersymmetric bosonic configuration described in section \ref{sec:E3bos}, we start from the $\kappa$-symmetric fermionic effective action for (Minkowskian) D-branes obtained in \cite{Marolf:2003vf, Martucci:2005rb}  and we Wick rotate it. The world-volume fermions are described by a Green-Schwarz-like ten-dimensional bispinor 
\be\label{Theta}
\Theta=\left(\begin{array}{c} \Theta_1 \\ \Theta_2 \end{array} \right)
\ee
In Minkowskian signature, $\Theta_1$ and $\Theta_2$ are MW in ten dimensions. After the Wick rotation, they lose their reality condition and are analytically continued to Weyl spinors whose precise form will be given in section \ref{sec:ferm}. Half of $\Theta$ is redundant because
of the $\kappa$-symmetry but, given that we are expanding around a supersymmetric configuration, this can be fixed in a natural way by imposing the constraint
\be\label{kappafixing}
\Gamma_{\rm E3}\Theta_2=-\Theta_1
\ee  
Taking into account (\ref{kappafixing}),  the fermionic effective action assumes the following general form
\be\label{fermact} S_{\rm F}=\ii\int_D\d^4\sigma\, e^{-\phi}\sqrt{\det \calm}
\, \Theta^T\calc\Big[(\calm_\sigma^{-1})^{AB}\Gamma_A
\cald_B-\frac12\calo\Big]\Theta
\ee
which constitutes the starting point of our analysis.

Let us explain (\ref{fermact}).  $\calc$ is the ten-dimensional complex conjugation matrix, which in our representation is equal to (Minkowskian) $\Gamma^{\ul 0}$. Furthermore, to unclutter  the formulas we have introduced  $\calm\equiv g|_D+\calf$, while $\calm_\sigma^{-1}$ is the inverse of $\calm_\sigma\equiv g|_D+\calf\,\sigma_3$, where Pauli matrices act on the bispinors of the form (\ref{Theta}). We remind the reader that $g|_D$ refers to the pull-back of the complete string frame metric. The operators $\cald_A$ and $\calo$ 
corresponds to the pull-back of those defining the (string frame) bulk gaugino and gravitino supersymmetry transformations:
\be\label{susy operators}
\begin{aligned}
\cald_M&=\left[\nabla_M+\frac14\iota_MH_{\it 3}\sigma_3+\frac{1}{16}e^\phi
\left(\begin{array}{cc} 0 & F \\ -\lambda(F)& 0 \end{array}\right)\Gamma_M\right]\ ,\cr
\calo &=\left[ \d\phi+\frac12\,H_{\it 3}\sigma_3+\frac{1}{16}e^{\phi}\,\Gamma^M  \left(\begin{array}{cc} 0 & F \\ -\lambda(F)& 0 \end{array}\right) \Gamma_M\right] \ .
\end{aligned}
\ee
where $\lambda$ acts on a ${\it k}$-form $\omega_{\it k}$ as $\lambda(\omega_{\it k})=(-)^{{\it k(k-1)}/2}\omega_{\it k}$ and all forms are implicitly contracted with gamma matrices, e.g.\ $H_{\it 3}\rightarrow \frac{1}{3!}H_{M_1M_2M_3}\Gamma^{M_1M_2M_3}$ or $\iota_MH_{\it 3}\rightarrow \frac{1}{2!}H_{MN_1N_2}\Gamma^{N_1N_2}$.

\subsection{World-volume fermions}
\label{sec:ferm}

Let us now come back to the fermions $\Theta$. As in \cite{Bianchi:2011qh} one can use the bulk and brane supersymmetric structure we are expanding about, to find a parametrization of $\Theta$ which automatically satisfies the constraint  (\ref{kappafixing}). 
First, we introduce complex coordinates $(s^a,\bar s^{\bar b})$, $a,\bar b=1,2$ on the E3 world-volume.  
Then, we write
\be
\Theta_1=\Theta^{\rm R}_1+\Theta^{\rm L}_1\qquad~~~~ \Theta_2=\Theta^{\rm R}_2+\Theta^{\rm L}_2
\ee
and set
\begin{subequations}\label{Fsplit}
\begin{align}
& \left\{ \begin{array}{l}
\Theta_1^{\rm L}=\frac{1}{2}e^{\frac{3A}{2} +\frac{\phi}{8}}\Big(\lambda\otimes \eta^*+(M^{-1})^{a\bar b}\psi_{\bar b}\otimes \gamma_a\eta+\frac14\rho_{ab}\otimes \gamma^{ab}\eta^*\Big) \\
\Theta_2^{\rm L}=\frac{\ii}{2} e^{\frac{3A}{2} +\frac{\phi}{8}}\Big(\lambda\otimes \eta^*-(M^{-1})^{\bar a b}\psi_{\bar a}\otimes \gamma_b\eta+\frac{1}4\rho_{ab}\otimes \gamma^{ab}\eta^*\Big) \end{array}\right. \\
&\left\{ \begin{array}{l} \Theta_1^{\rm R}=\frac{\ii}2e^{\frac{3A}{2} +\frac{\phi}{8}}\Big(\tilde\lambda\otimes \eta+(M^{-1})^{ \bar a b}\tilde\psi_b\otimes \gamma_{\bar a}\eta^*+\frac14\tilde\rho_{\bar a\bar b}\otimes \gamma^{\bar a\bar b}\eta \Big)
\\
\Theta_2^{\rm R}=-\frac{1}{2} e^{\frac{3A}{2} +\frac{\phi}{8}} \Big(\tilde\lambda\otimes \eta-(M^{-1})^{a\bar b}\tilde\psi_a\otimes \gamma_{\bar b}\eta^*+\frac{1}4\tilde\rho_{\bar a \bar b}\otimes \gamma^{\bar a \bar b}\eta\Big)
\end{array}  \right.
\end{align}
\end{subequations}
where  $\eta$ is the bulk spinor of (\ref{killing}), the overall appearance of warping and dilaton is  for later convenience and  we have introduced 
\be
M\equiv e^{2A-\frac\phi2}\calm\equiv h+e^{2A-\frac\phi2}\calf
\ee
Notice that, being the brane holomorphically embedded and $\calf$ purely (1,1), $M$ has only $(1,1)$ components as well. 

In the decomposition (\ref{Fsplit}) the independent degrees of freedoms are now parametrized by world-volume fermions $\lambda,\psi,\rho,\tilde\lambda,\tilde\psi,\tilde\rho$, which are forms rather than spinors from the four-dimensional world-volume viewpoint.  In other words, the natural parametrization (\ref{Fsplit}) explicitly realizes the `topological twist' which is expected to be a generic feature of branes on curved manifolds \cite{Bershadsky:1995qy}. On the other hand, the new world-volume fermions still carry a spinorial index along the external $\mathbb{R}^4$ directions. More precisely $\lambda,\psi,\rho$ are left-handed ($\hat\gamma_5=-1$) while $\tilde\lambda,\tilde\psi,\tilde\rho$ are right-handed ($\hat\gamma_5=+1$). Hence, in two-index notation for four-dimensional Weyl spinors, we can summarize the fermionic field content more explicitly as follows:
\be\label{fermcont}
\begin{aligned}
&\lambda^\alpha\quad~~~~~ \psi^\alpha\equiv \psi^\alpha_{\bar a}\,\d \bar s^{\bar a}\quad~~~~~ \rho^\alpha\equiv \frac12 \rho^\alpha_{ab}\,\d s^{a}\wedge \d s^b\qquad~~~\text{(left-handed)}\\
& \tilde\lambda^{\dot\alpha}\quad~~~~~\tilde\psi^{\dot\alpha}\equiv \psi^{\dot\alpha}_{a}\,\d  s^{a}\quad~~~~~ \tilde\rho^{\dot\alpha}\equiv \frac12 \rho^{\dot \alpha}_{\bar a\bar b}\,\d \bar s^{\bar a}\wedge \d \bar s^{\bar b}\qquad~~~\text{(right-handed)}
\end{aligned}
\ee
Here we see more precisely the fate of world-volume fermions under Wick rotation we alluded to above. 
In a Minkowskian brane, the left- and right-handed sector would be related by complex conjugation and one would count real degrees of freedom. On E3-branes this is no longer the case and the left- and right-moving fermions appearing in (\ref{fermcont}) should be considered as complex fields each counting as one independent degree of freedom. 

 One of the virtues of the decomposition \ref{Fsplit} is that the fermions (\ref{fermcont}) transform with a well defined U(1)$_Q$ charge $q_Q$ 
under the type IIB $\text{SL}(2,\mathbb{Z})$ duality group -- see section \ref{sec:bulkduality}. This is explained in \cite{Bianchi:2011qh} and the argument given therein, being independent on the presence of bulk fluxes, works for our case as well. Let us then summarize the properties of the world-volume spinors by the following table:
\be
\label{fermcharges}
\begin{array}{|c|c|c|} \text{fermions}& q_Q &\text{associated bundle}\\ \hline
\lambda^\alpha&  0 &  S_-\otimes \Lambda_D^{0,0}\\
\psi^\alpha & 0 & S_-\otimes \Lambda_D^{0,1}\\
\rho^\alpha & 0 & S_-\otimes \Lambda_D^{2,0}\\
\tilde\lambda^{\dot\alpha}&  -1 &S_+\otimes \Lambda_D^{0,0}\otimes L_Q^{-1}\\
\tilde\psi^{\dot\alpha} & +1 &S_+\otimes \Lambda_D^{1,0}\otimes L_Q\\
\tilde\rho^{\dot\alpha} & -1 & S_+\otimes \Lambda_D^{0,2}\otimes L_Q^{-1}\end{array}
\ee
Here $S_{\pm}$ are the right/left handed spin bundles along $\mathbb{R}^4$, $\Lambda_D^{p,q}$ refers to the bundle of $(p,q)$-forms on $D$ and the restriction to the E3-brane of the bundles defined in the bulk is understood.

\subsection{Topologically twisted fermionic effective action}

The next step is to plug the decomposition (\ref{Fsplit}) into the general formula (\ref{fermact}) and obtain an action for the new fields (\ref{fermcont}).  Although straightforward, the explicit calculation turns out to be quite lengthy and to require a certain amount of ingenuity in order to simplify the final result.  We address the reader interested in the technical details of the derivation to appendix \ref{app:details} and here we just quote the final result. 

In its final form, it is convenient to split the fermionic E3-brane action in two parts, one for the left-handed sector and the other one for the right-handed sector:
\be\label{action1}
S^{\rm ferm}=S^{\rm ferm}_{\rm L}+S^{\rm ferm}_{\rm R}
\ee
which we presently discuss one-by-one in some detail. Recall that we are focusing on the case of supersymmetric backgrounds. The possible contribution from a supersymmetry breaking flux $G_{0,3}\neq 0$ is discussed in section \ref{sec:SBsub}. 

Let us start with the left-moving part, which reads
\be
\begin{aligned}\label{FactionL}
S^{\rm ferm}_{\rm L}=&2\int_D\left( *\psi\wedge\del^+\lambda -*\rho\wedge\delbar^-\psi \right)+\frac12\int_De^{2A}\,*\rho\wedge \cals\cdot\rho\\
\end{aligned}
\ee
Let us explain the notation in some detail. $*$ is the Hodge star computed by using the induced world-volume K\"ahler metric $h_{a\bar b}$ and the suffixes ${}^\pm$ indicate a modification due to the warping, namely $\del^\pm\equiv\del\pm  \del A\equiv e^{\mp A}\,\del\, e^{\pm A}$. We always keep implicit the contraction of the external four-dimensional spinorial indices by the four-dimensional charge conjugation matrix $C=\ii\gamma^{\ul 0}$. For instance:
\be
\psi(\ldots)\lambda\equiv \psi^TC(\ldots)\lambda\equiv C_{\alpha\beta}\psi^\alpha(\ldots)\lambda^\beta\equiv \psi^\alpha(\ldots)\lambda_\alpha
\ee
The effect of fluxes is encoded in the mass-like operator  $\cals:\Lambda_D^{2,0}\rightarrow \Lambda_D^{0,2}$,  which acts as $\cals\cdot \rho=\frac{1}{2!2!}\cals_{\bar a\bar b}{}^{cd}\rho_{cd}\,\d \bar s^{\bar a}\wedge\d\bar s^{\bar b}$ and  is defined as follows in terms of the bulk and world-volume fluxes and extrinsic curvature $\calk$ (computed from the bulk K\"ahler metric):
\be\label{S1}
\cals_{\bar a\bar b}{}^{cd} \equiv -\,\Big[e^{-\phi} \calk^{\bar{\bf v}}{}_{\bar e[\bar a}\calf_{\bar b]}{}^{\bar e} -\frac{\ii}{4}\, \bar{G}^{\bar{\bf v}}{}_{\bar a\bar b}\Big]\,\bar\Omega_{\bar{\bf v}}{}^{cd}
\ee
Here an in the following, world-volume indices are lowered and raised by using the world-volume metric $h_{a\bar b}$.

Notice that  all world-volume flux non-linearities which where present in the original GS-like action (\ref{fermact})  have remarkably disappeared in the left-handed action (\ref{FactionL}). Furthermore, the non-trivial bulk and world-volume fluxes eventually produce a mass-like term only for the $\rho$-fields.

The story for the right-handed sector is different. 
Indeed, $S^{\rm ferm}_{\rm R}$ is slightly more complicated than $S^{\rm ferm}_{\rm L}$:
\be
\begin{aligned}\label{FactionR}
S^{\rm ferm}_{\rm R}=&2\int_D\left( *\tilde\psi\wedge\delbar_Q^-\tilde\lambda -*\tilde\rho\wedge\del_Q^+\tilde\psi \right)\\
&+\frac12\int_D\,e^{2A}\,\left(*\tilde\rho\wedge\tilde\cals\cdot\tilde\rho+2\,e^{2A}\,*\tilde\psi\wedge\tilde\calu\cdot \tilde\rho+*\tilde\psi\wedge\tilde\calr\cdot\tilde\psi
\right)
\end{aligned}
\ee
Let us again explain the notation in detail.
We are using the U(1)$_Q$ covariant (anti)-holomorphic exterior derivatives defined by $\del_Q=\del-\ii q_Q Q_{1,0}$ and $\delbar_Q=\delbar-\ii q_Q Q_{0,1}$, with  charges $q_Q$ as listed in (\ref{fermcharges}), and the suffixes $^\pm$ are used as above.  (Since we are working on the world-volume all bulk quantities, as for instance $Q_{\it 1}$, are implicitly pulled-back to $D$.)
As for the left-handed sector,  we have introduced the operators 
\be
\tilde\cals:\Lambda_D^{0,2}\otimes L^{-1}_Q\rightarrow \Lambda_D^{2,0}\otimes L_Q\quad~~~~~\tilde\calu:\Lambda^{0,2}_D\rightarrow\Lambda^{0,1}_D\quad~~~~~\tilde\calr:\Lambda_D^{1,0}\otimes L_Q\rightarrow \Lambda_D^{0,1}\otimes L^{-1}_Q
\ee
which play the role of mass terms and act more explicitly as follows
\be
\tilde\cals\cdot\tilde\rho=\frac{1}{2!2!}\tilde\cals_{ab}{}^{\bar c\bar d}\, \tilde\rho_{\bar c\bar d} \d s^a\wedge\d s^b\quad~~~~~~ 
\tilde\calu\cdot\tilde\rho=\tilde\calu{}^{\bar a}\, \tilde\rho_{\bar a\bar b} \,\d \bar s^{\bar b}\quad~~~~~~ 
\tilde\calr\cdot\tilde\psi=\tilde\calr_{\bar a}{}^b\tilde\psi_b\,\d \bar s^{\bar a}
\ee
where
\begin{subequations}\label{tilded operators}
\begin{align}
\tilde\cals_{ab}{}^{\bar c\bar d} &\equiv -e^{-\phi}\, \Big[\calk^{\bf v}{}_{e[a}\calf_{b]}{}^e +\frac{\ii}{4}\, e^{4A}(\calf\cdot\calf)\,G^{\bf v}{}_{ab}\Big]\,\Omega_{{\bf v}}{}^{\bar c\bar d}\label{S2}\\
\tilde\calu^{\bar a}&\equiv\ii\,\calf_{b\bar c}\,G^{b\bar c\bar a}\label{calu}\\
\tilde\calr_{\bar a}{}^b&\equiv\,\bar\Omega^{{\bf v}c}{}_{\bar a}\,\Big[ {\bf v}(\phi)e^{-\phi}\calf_c{}^b+ \frac{\ii}{2}G_{{\bf v}c}{}^b]+\tilde\calr'_{\bar a}{}^b\label{calr}
\end{align}
\end{subequations}
with
\be\label{calrprime}
\tilde\calr'_{\bar a}{}^b\equiv e^{4A-\phi}\,\bar\Omega^{{\bf v}c}{}_{\bar a}\,\calf_c{}^b\, 
 \Big[ {\bf v}(e^{-4A})-\frac12{\bf v}(\phi)e^{-\phi}\calf\cdot\calf-\frac{\ii}{2}\,\calf\cdot \iota_{\bf v}G_{\it 3}  \Big]\,\sqrt{\frac{\det h}{\det M}}
\ee
Here ${\bf v}$ refers to a $(1,0)$ direction orthogonal to $D$. In other words, ${\bf v}$ is a  section of the $(1,0)$ normal bundle $N^{1,0}_D$.
Then, as usual, ${\bf v}(\phi)\equiv {\bf v}^m\partial_m\phi$.

From (\ref{FactionR}) we see that fluxes produce not only a mass term for $\tilde\rho$, analogous to that for $\rho $ in (\ref{FactionL}), but also
a mass term for $\tilde\psi$ and an off-diagonal term mixing $\tilde\psi$ and $\tilde\rho$. Notice that the mass operator  $\tilde\calr'$ provides the only non-polynomial contribution to the fermionic action. We will come back to this term in the following.

\subsection{Inclusion of supersymmetry breaking fluxes}
\label{sec:SBsub}

So far we have assumed that the bulk configuration is supersymmetric. Indeed, we have heavily used the supersymmetry conditions of the E3-brane, which strictly speaking make sense only for supersymmetric backgrounds.

Nevertheless, a supersymmetry breaking $G_{0,3}$ flux  does not spoil 
the nice features of the underlying geometrical structures. In particular there is still a bulk K\"ahler structure and then it is still sensible to require that an E3-brane wraps a divisor with a $(1,1)$ and primitive world-volume flux. Even though these conditions do not anymore imply that the brane is supersymmetric, they are still naturally selected by their good properties. So, we can keep these conditions even in the case with $G_{0,3}\neq 0$ and compute the effective fermionic action as we have done in the supersymmetric case. As a result, the action (\ref{action1}) must be supplemented by the following  additional contribution:
\be\label{SBaction}
\begin{aligned}
\Delta^{\rm SB} S^{\rm ferm}&\equiv\ \Delta^{\rm SB} S_{\rm L}^{\rm ferm}+\Delta^{\rm SB} S_{\rm R}^{\rm ferm}\cr
 &=\frac{\ii}8\int_D\ e^{2A}\,(\bar{G_{\it 3}}\cdot\bar\Omega)\,\lambda\lambda\, *1+\frac{\ii}{16}\int  e^{6A-\phi}\,\calf\wedge\calf\,
	(G_{\it 3}\cdot\Omega)\,\tilde\lambda\tilde\lambda
\end{aligned}
\ee

To summarize, on a general (possibly non supersymmetric) flux  F-theory background, 
the {\em complete} quadratic fermionic effective action governing  fermions living on  an E3-brane wrapping a holomorphic four-cycle and supporting a possible $(1,1)$ and primitive world-volume flux $\calf$, is given by  
\be\label{fullaction}
S^{\rm ferm}_{\rm tot}=S_{\rm L}^{\rm ferm}+S_{\rm R}^{\rm ferm}+\Delta^{\rm SB} S_{\rm }^{\rm ferm}
\ee
where $S_{\rm L}^{\rm ferm}$, $S_{\rm R}^{\rm ferm}$  and $\Delta^{\rm SB} S_{\rm }^{\rm ferm}$ are given in (\ref{FactionL}), (\ref{FactionR})
and (\ref{SBaction}) respectively, in terms of the mass operators $\cals$, $\tilde\cals$, $\tilde\calu$ and $\tilde\calr$ defined in (\ref{S1}), (\ref{S2}), (\ref{calu}) and (\ref{calr}). 

This fermionic action constitutes one of our main results. 
As a final comment before turning to the inspection of its properties and implications, notice that the actions (\ref{fermact}) and (\ref{action1}) are not strictly identical,  via the field redefinition (\ref{Fsplit}). Indeed, they differ by some terms which vanish by  imposing  the world-volume Bianchi identity $\d\calf=H|_D$. These additional terms can be found in equation (\ref{BI surviving term}) and are discussed thereafter.

\section{O-planes and ${\rm SL}(2,\mathbb{Z})$ duality}
\label{sec:actionduality}

As we have already stressed,  F-theory backgrounds are characterized by non-trivial  ${\rm SL}(2,\mathbb{Z})$ dualities which are part of the vacuum configuration. In certain cases, these codify the non-perturbative resolution of $\Omega$7-planes in weakly coupled CY compactifications. Furthermore, $\Omega$3-planes can appear as well. In particular, in the $\Omega$-plane description one can employ complementary perturbative string theory methods in order to go beyond the supergravity approximation. In this section we describe how the dualities and orientifold projections are accommodated by  the action (\ref{action1}). 

\subsection{Orientifold projections}
\label{sec:Oplanes}

In the orientifold limit, one can describe the background in terms of a covering CY space  $\tilde X$. The $\Omega$-planes sit at the fixed loci of space involutions $\sigma:\tilde X\rightarrow \tilde X$, with $\sigma\circ\sigma=\text{Id}$, and the actual compactification space is given by $X=\tilde X/\sigma$. In particular, the $\Omega$-planes compatible with the backgrounds we are discussing are $\Omega$7-planes and $\Omega$3-planes, each associated with its own geometrical involution, $\sigma_{\Omega 7}$
and $\sigma_{\Omega 3}$.  

The orientifold projection selects the string states which are invariant under an involution which combines the action of $\sigma$ with  world-sheet parity. Hence, all the supergravity fields characterizing the background must be invariant under such a projection.
For $\Omega$3/$\Omega$7 planes this requirement boils down to the following conditions:
\be
\sigma^*g=g\qquad \sigma^*\phi=\phi\qquad \sigma^*G_{\it 3}=-G_{\it 3}\qquad \sigma^* F_{\it 5}=F_{\it 5}
\ee
Furthermore, the CY structures $J$ and $\Omega$ on $\tilde X$ must fulfill the projections
\be\label{SU(3)proj}
\sigma^* J=J\qquad \sigma^*\Omega=-\Omega
\ee
This can be understood by noticing that the orientifold projection act on the bulk ten-dimensional supersymmetry generators $(\epsilon_1,\epsilon_2)$ as $\sigma^*\epsilon_2=\epsilon_1$, which for our backgrounds translates into the condition $\sigma^*\eta=\ii\eta$ on the internal spinor. Hence, the projections in (\ref{SU(3)proj}) follow from (\ref{SU(3)def}).

Let us now pass to the E3-brane. It wraps a divisor $D\subset X$, which uplifts to a covering divisor $\tilde D\subset \tilde X$.  
Clearly, $\tilde D$ must be invariant under the orientifold involution $\sigma$, {\em i.e.}\ $\sigma(\tilde D)\equiv \tilde D$.
On the other hand, the world-volume flux $\calf$  on $\tilde D$ must satisfy the following projection condition
\be\label{fluxproj}
\sigma^*\calf=-\calf
\ee

The action of the $\Omega$3/$\Omega$7 involution on the fermions supported by the E3-brane can be obtained by requiring that the Green-Schwarz fermions $(\Theta_1,\Theta_2)$ living on $\tilde D$ are projected as $(\epsilon_1,\epsilon_2)$.\footnote{At least, up to a $\kappa$-symmetry transformation which in our case turns out to be vanishing.} Indeed, taking into account (\ref{fluxproj}) and the bulk condition $\sigma^*\eta= \ii\eta$, we see that the ansatz (\ref{Fsplit}) satisfies such projection condition provided that
\be
\begin{aligned}
&\sigma^*\lambda=\lambda\qquad\qquad~~\sigma^*\psi=\psi\qquad\qquad~~~ \sigma^*\rho=\rho\cr
&\sigma^*\tilde\lambda=-\tilde\lambda\qquad\qquad\sigma^*\tilde\psi=-\tilde\psi\qquad\qquad \sigma^*\tilde\rho=-\tilde\rho
\end{aligned}
\ee 

It is easy to check that the action (\ref{fullaction}) is invariant under the orientifold involution, as it should.

\subsection{${\rm SL}(2,\mathbb{Z})$ duality}
\label{sec:SL2Z}

Let us now come back to a complete F-theory background, in which possible $\Omega$7-planes are resolved in bound states of $(p,q)$ 7-branes. In order to simplify the discussion, we assume that no $\Omega$3-planes are present.  These could be easily introduced by applying the discussion of section \ref{sec:Oplanes} just to the $\Omega$3-plane case, since the possible $\Omega$7-planes are already incorporated in the bulk solution. 

In order for the action (\ref{action1}) to be well defined on general F-theory vacua, we have to check that it behaves properly under ${\rm SL}(2,\mathbb{Z})$ duality transformations. We have already discussed how the different fields transform under ${\rm SL}(2,\mathbb{Z})$. In particular, all the fields appearing in (\ref{action1}) transform with a definite U(1)$_Q$ charge $q_Q$ (as defined in section \ref{sec:bulkduality}). The U(1)$_Q$ charges of the world-volume fermions are listed in (\ref{fermcharges}). For the reader's convenience, let us here summarize how the relevant bosonic fields transform under ${\rm SL}(2,\mathbb{Z})$: the metric $h$ and the warping $e^{A}$ are neutral, \ie have U(1)$_Q$ charge $q_Q=0$, while
\be
\label{boscharges}
\begin{array}{|c|c|c|} \text{fields}& q_Q &\text{associated bundle}\\ \hline
e^{-\frac{\phi}{2}}\calf & 1 & \Lambda^{1,1}_D\otimes L_Q\\
e^{\frac{\phi}{2}}G_{\it 3}&  -1 &  (\Lambda^{2,1}_X\oplus \Lambda^{0,3}_X)\otimes L^{-1}_Q\\
e^{-\frac{\phi}{2}}\Omega & 1 &  \Lambda^{3,0}_X\otimes L_Q\\
\del\phi=\frac{\ii}{2}e^\phi\del\tau & -2 &\Lambda^{1,0}_X\otimes L^{-2}_Q
\end{array}
\ee

By using the information of (\ref{fermcharges}) and (\ref{boscharges}) we can now inspect the properties of our complete action (\ref{fullaction}) under ${\rm SL}(2,\mathbb{Z})$. In particular, it is easy to verify that $S^{\rm ferm}_{\rm L}$ and $\Delta^{\rm SB}S^{\rm ferm}$
 are manifestly {\em invariant} under ${\rm SL}(2,\mathbb{Z})$. This is an important property, which appears to be necessary for the action to be well formulated. Indeed this property played an important role in the arguments  of \cite{Bianchi:2011qh}, in the contest of F-theory vacua without bulk fluxes and warping.

 On the other hand life is not so simple for the right-handed sector.  Indeed, already at a first look, it is clear that not all the terms contained in $S^{\rm ferm}_{\rm R}$ -- see Eq.~(\ref{FactionR}) -- are invariant under ${\rm SL}(2,\mathbb{Z})$. Actually, almost all of them are invariant, with the exclusion of  the contribution $\tilde\calr'$ to the operator $\tilde\calr$, defined in (\ref{calrprime}) and (\ref{calr}) respectively.   

 This could seem to pose an interpretational problem for our effective action, at least for the term involving
 $\tilde\calr'$. However, as we discuss in the following section, things are not so bad as they could appear.

\section{Off-shellness and SL(2,$\mathbb{Z}$) invariance}
\label{sec:offshell}

The problem raised at the end of the previous section is closely related to the following crucial observation: because of the background and world-volume fluxes, the supersymmetric E3-brane configurations we are discussing about {\em do not} extremize the E3-brane effective action.

 In order to show this, one should compute the first order variation of the bosonic effective action (\ref{action1}) around a supersymmetric configuration. This is explicitly done in appendix \ref{app:eom}. Consider a generic variation of the embedding described by a section $\delta v=\delta \varphi\,{\bf v}+\delta\bar\varphi\,\bar{\bf v}$ of the normal bundle $N_D=N^{1,0}_D\oplus N^{0,1}_D$, which we have decomposed into its (1,0) and (0,1) components by using the (1,0) normal vector ${\bf v}$ and its complex conjugate $\bar{\bf v}$. Then, one can check that $\delta_{\delta \bar\varphi}S_{\rm E3}=0$ but also that, on the other hand, $\delta_{\delta \varphi}S_{\rm E3}$ does not identically vanish:
\be\label{offshell}
\delta_{\delta \varphi}S_{\rm E3}=4\pi \int_{D}\d^4\sigma\sqrt{\det h}\,\delta\varphi\Big[{\bf v}(e^{-4A})-\frac12{\bf v}(\phi)e^{-\phi}\, \calf\cdot \calf-\frac{\ii}{2}\,\calf\cdot \iota_{\bf v} G_{\it 3}\Big]
\ee 
Another non-vanishing contribution is obtained if we try to vary the action with respect to the gauge field, see equation (\ref{Avar}). These problems are related to the fact that the action (\ref{E3action1}) is complex and the non-vanishing bulk R-R fluxes make the  CS term contribute non-trivially to the equations of motion.\footnote{Furthermore, the R-R potentials are only locally defined and, on top of it, there are possible sources for the R-R fluxes, which make the definition of R-R potentials, and then of the CS term itself, even more subtle. In this respect, the situation is similar to what happens for world-sheet instantons in heterotic compactifications, in which the CS-like coupling $\int B_{\it 2}$ must be properly interpreted since $H_{\it 3}$ is not closed but must obey the Bianchi identity  $\d H_{\it 3}\sim (\tr R\wedge R-\tr F\wedge F)$. The proper way to handle this issue has been explained in \cite{Witten:1999eg} and requires to consider at the same time the factor $e^{2\pi \ii\int B_{\it 2}}$ and the fermionic Pfaffian in the path integral. We expect a similar argument, appropriately generalised, to be valid in our case as well, but a proper understanding of this important problem is beyond the scope of the present paper.}

Having an off-shell configuration could appear inconsistent at first sight. Indeed, to the best of our knowledge this fact has not  been fully appreciated so far in the the literature on instantonic D-branes, mainly because it is associated to the presence of fluxes, which are not often considered. However, this effect is quite common in standard instanton calculus. Already in the seminal paper \cite{'tHooft:1976fv} it was observed that in Euclidean space it is not compulsory  to start a semiclassical computation from a classical configuration which extremizes the action. In our case it is the requirement of supersymmetry  which selects the preferred brane configurations, the ones  which can contribute to the F-terms of the low-energy effective theory, even though they can violate the equations of motion. This violation is small in the large universal modulus regime, further discussed in section \ref{sec:pert}, and in this sense our E3-brane configurations can be seen as approximate solutions of the classical equations. Moreover, like in standard instanton calculus, the effective four-dimensional F-terms generated by the E3-brane instanton may turn out to be independent of the world-volume sector which is responsible for the violation of the equations of motion. 

On the other hand, it is easy to see how the off-shellness of the supersymmetric E3-brane instanton is related to the breaking of manifest SL(2,$\mathbb{Z}$) invariance of the right-handed fermionic action. Indeed, the mass operator $\tilde\calr'$ defined in (\ref{calrprime}) is exactly proportional to the integrand appearing on the right-hand side of (\ref{offshell}). More precisely, by using (\ref{offshell}) we can write
\be
\tilde\calr'_{\bar a}{}^b=\,\frac{1}{4\pi}e^{4A-\phi}\bar\Omega^{{\bf v}c}{}_{\bar a}\, \calf_c{}^b \,\sqrt{\frac{\det h}{\det M}}\, \frac{\delta S^{\rm susy}_{\rm E3}}{\delta \varphi}
\ee
Clearly, $\tilde\calr'_{\bar a}{}^b\neq 0$ is non-vanishing because ${\delta S^{\rm susy}_{\rm E3}}/{\delta \varphi}$ is non-vanishing. 

In fact, this observation is in agreement with what is expected from the dual M-theory picture, which will be discussed more in detail in section \ref{sec:uplift}. The E3-brane instanton is dual to a Euclidean M5-brane wrapping a `vertical' divisor in an elliptically fibered CY four-fold, \ie a divisor which wraps the elliptic fiber $T^2_\tau$ (which has complex structure $\tau$). In the M-theory picture the type IIB SL(2,$\mathbb{Z}$) duality is geometrized into the modular group of large diffeomorphisms of $T^2_\tau$. Hence, if it was possible to write a diffeomorphism invariant quadratic fermionic action on the  Euclidean M5-brane, this would correspond to a SL(2,$\mathbb{Z}$)-duality invariant quadratic fermionic action on the M5-brane and would then contradict our result. But it is known that this is not possible in general. Indeed, the M5-brane diffeomorphism invariant action obtained in  \cite{Kallosh:2005yu} crucially uses the fact that the bosonic M5-brane configuration solves the bosonic equations of motion.\footnote{We thank Dmitri Sorokin for clarifying discussions on this point.}
This is perfectly in agreement with our result: the violation of the SL(2,$\mathbb{Z}$) invariance  is proportional to (part of) the bosonic equations of motion which are not satisfied by our E3-brane configuration. 

Of course, having a manifest SL(2,$\mathbb{Z}$) invariance of the action could be very useful in order to study the E3-brane path-integral. In order to achieve it in our setting, one should probably reformulate our result by introducing an auxiliary field as the one used in \cite{Pasti:1997gx,Bandos:1997ui}.  On the other hand, we stress that the violation of the manifest  SL(2,$\mathbb{Z}$) invariance in our  E3-brane fermionic effective action is, remarkably, restricted just to the term  containing $\tilde\calr'$, which furthermore  is subleading in a sense that will be made precise in  section \ref{sec:pert}. Hence, as we will see, our action can in fact provide all the necessary information for understanding the impact of fluxes on the E3-brane generated non-perturbative effects.


\section{Perturbative expansion}
\label{sec:pert}

In order to make the right-handed effective action look more tractable, we will  work in a perturbative regime, in which all the terms can be organised in a hierarchical expansion.  The natural perturbative parameter  is the  overall breathing modulus, which can be taken as the parameter of  the supergravity approximation scheme, which assumes large internal volumes compared to the string length scale. Of course, experience tells us that supersymmetry often comes to rescue and assures that the supergravity effective theory provides sensible information even well beyond the leading order approximation. Nevertheless, a priori, the effective action (\ref{fermact}) we started from is supposed to be valid only in the long-wavelength approximation.

The bulk breathing mode can be identified with the integration constant $c$ appearing in (\ref{breathing}). Of course, there is still an ambiguity in  the split of $e^{-4A}$ into $c+e^{-4\hat A}$, but this can be fixed by imposing any normalization condition on $e^{-4\hat A}$. For instance for our purposes a natural choice would be  
\be
\int_D \d^4 \sigma\,e^{-4\hat A}\sqrt{\det h}=1
\ee
Hence, the large breathing mode limit corresponds to the regime in which
\be
\varepsilon\equiv \frac{1}{\sqrt{c}}
\ee 
is very small. At tree level, $\varepsilon$ can be see considered as a freely adjustable parameter. Hence,  it is legitimate to choose $\varepsilon\ll 1$ and reorganize our fermionic action (\ref{fullaction}) as an expansion in $\varepsilon$.  As we will see, such an expansion can be physically very useful.  

By using the formula $e^{2A}\simeq \varepsilon[1-\frac{1}{2}\,e^{-4\hat A}\varepsilon^2 +\calo(\varepsilon^4)]$ and $\d A\simeq \varepsilon^2\d \hat A[ e^{-4\hat A}-\varepsilon^2e^{-8\hat A}+\calo(\varepsilon^4)] $, one can immediately work out the expansion of all the terms in (\ref{fullaction}). However it is useful to note that, at the practical level, the large breathing mode expansion of our fermionic action can be identified with an expansion in powers of the warping $e^{2A}$ (up subleading corrections).
Furthermore, as one can readily check by looking at our fermionic effective action,  this is in turn equivalent to expanding it in powers of the bulk and world-volume fluxes, since each one comes with a factor $e^{2A}$. This observation has a simple physical explanation. Taking a larger breathing mode has the effect of diluting the fluxes, which can then be considered as expansion parameters. 

We can now easily realize that the kinetic terms in the effective action are of order $\calo(1)$, the mass terms containing $e^{2A}\cals$, $e^{2A}\tilde\cals$ and $e^{2A}\tilde\calr$ are of order $\calo(\varepsilon)$ and the mass term containing $e^{4A}\tilde\calu$ is of order $\calo(\varepsilon^2)$. In particular, the contribution from $e^{2A}\tilde\calr'$ to $e^{2A}\tilde\calr$ is of order $\calo(\varepsilon^3)$ and can then be considered as subleading. Hence,  by working in this perturbative regime all types of terms have a leading order contribution which is manifestly SL(2,$\mathbb{Z}$) invariant and  then, pragmatically speaking, the  SL(2,$\mathbb{Z}$) issue can be ignored.

We then conclude that the issues on off-shellness and on SL(2,$\mathbb{Z}$) invariance for the right-handed sector  are strongly related but also that they are harmless up to order $\calo(\varepsilon^2)$.  The inclusion of higher order terms would  require the development of new techniques for off-shell brane instantons which we leave to the future.

\section{Fermionic zero modes}
\label{sec:zm}

Zero modes play a distinguished role in the evaluation of instanton induced effective interactions  and in this section we would like to study how they are affected by the fluxes.  
For the sake of clarity, we consider first the case of a supersymmetric background, postponing the inclusion of a non-vanishing $G_{0,3}$ flux to section \ref{sec:SBmodes}. 

\subsection{Left-handed zero modes}
In the absence of supersymmetry breaking $G_{0,3}$-flux, the effective action for the left-handed fermionic sector is given by (\ref{FactionL}). It can be written in the form  
 \be
 \frac12\int_D\d^4\sigma\sqrt{\det h}\,\Psi \calt_{\rm F}\Psi
 \ee 
 where
 \be
 \Psi\equiv\left(\begin{array}{c} \lambda \\ \psi \\ \rho\end{array}\right)\in S_-\otimes (\Lambda^{0,0}_D\oplus\Lambda_D^{0,1}\oplus \Lambda^{2,0}_D)
 \ee 
 and  $\calt_{\rm F}$ is the symmetric (but not hermitian) operator:
\be\label{calt}
\calt_{\rm F}=2\left(\begin{array}{ccc} 
0 & (\delbar^-)^\dagger & 0 \\
\del^+ & 0 & -(\del^+)^\dagger \\
0 & -\delbar^- & e^{2A}\cals
\end{array}\right)
\ee 
where $\cals:\Lambda_D^{2,0}\rightarrow\Lambda_D^{0,2}$ is the operator defined in (\ref{S1}).

Zero modes are given by the solutions of
\be\label{eqzeromodesL}
\calt_{\rm F}\Psi^{\rm z.m.}=0
\ee
Notice that by performing the warping-dependent rescalings
\be\label{warprescal} 
\lambda\rightarrow e^{-A}\lambda\quad~~~~\psi\rightarrow e^{A}\psi\quad~~~~\rho\rightarrow e^{-A}\rho
\ee
 the zero mode equations (\ref{eqzeromodesL}) can be rewritten in the following   warping independent form:
\begin{subequations}\label{Lzm}
\begin{align}
\delbar^\dagger\psi^{\rm z.m.}&=0\label{Lzm1}\\
\del\lambda^{\rm z.m.}-\del^\dagger\rho^{\rm z.m.}&=0\label{Lzm2}\\
\delbar\psi^{\rm z.m.}&=\frac12\,\cals\cdot\rho^{\rm z.m.}\label{Lzm3}
\end{align}
\end{subequations}
The condition (\ref{Lzm2}) is equivalent to separately imposing $\del\lambda^{\rm z.m.}=0$ and $\del^\dagger\rho^{\rm z.m.}=0$, then $\lambda^{\rm z.m.}$ and  $\rho^{\rm z.m.}$ must be harmonic (0,0) and (2,0) forms respectively.  On the other hand
(\ref{Lzm3}) admits a solution only if $\cals\cdot\rho^{\rm z.m.}$ is exact in $\delbar$-cohomology, which means that we can set $\cals\cdot\rho^{\rm z.m.}=2\delbar\alpha$, for some (0,1)-form $\alpha$ which, furthermore, can  be taken to be $\delbar^\dagger$-exact. Then, we can pick $\psi^{\rm z.m.}=\psi^{\rm z.m.}_{\rm harm}+\alpha$, with $\psi^{\rm z.m.}_{\rm harm}$ a harmonic $(0,1)$-form,  so that (\ref{Lzm1}) is fulfilled too. If, on the other hand, $\cals\cdot\rho^{\rm z.m.}$ is {\em not} $\delbar$-exact, then $\rho^{\rm z.m.}$ is lifted from the spectrum of zero modes.

Hence, the spectrum of left-handed zero modes can be summarized as follows.
If we denote by $\calh_{\delbar}$  the projector on the space of $\delbar$ harmonic forms on $D$ and by
\be\label{calp}
\calp_{2,0}\equiv \calh_{\delbar}\circ \cals\ :\  \text{Harm}_\del^{2,0}(D)\rightarrow \text{Harm}_{\delbar}^{0,2}(D)
\ee
the operators resulting from the composition of $\calh_{\delbar}$ and the mass operator $\cals$, then we have the following characterization of the spectrum
\be\label{lhzm}
\Psi^{\rm z.m.}\quad\Leftrightarrow\quad \text{Harm}_\del^{0,0}(D)\ \oplus\ \text{Harm}_{\delbar}^{0,1}(D)\ \oplus\  \ker\calp_{2,0} 
\ee
Notice that, by construction, the structure of this spectrum does not depend at all on the warping, assumed to be smooth and bounded, and in particular on the breathing mode, although the specific form of the zero modes depends on the warping through the rescaling (\ref{warprescal}).

We see that the zero modes $\lambda^{\rm z.m.}$ and $\psi^{\rm z.m.}$ are not sensitive to fluxes, but only to the underlying F-theory complex structure. For $\lambda^{\rm z.m.}$, which are just constant functions on $D$, this is in fact expected. Indeed $\lambda^{\rm z.m.}$'s are universal zero modes, usually denoted by $\theta^\alpha$. They are the goldstini of the broken supersymmetry $\zeta_L$, which acts non-linearly on $\lambda$ by $\delta_{\rm L}\lambda^\alpha\sim \zeta^\alpha_{\rm L}$.

\subsection{Right-handed zero modes}

Let us now pass to the right-moving sector, still assuming a supersymmetric bulk. It is governed by the effective action (\ref{FactionR}).
As we have discussed in section \ref{sec:SL2Z}, the massive term in $\tilde\psi\tilde\psi$ containing $\tilde\calr'$ breaks the manifest SL(2,$\mathbb{Z}$) symmetry of the fermionic action. On the other hand, in section \ref{sec:pert} we pointed out that, in a large breathing mode expansion,  this effect shows up only at order $\calo(\varepsilon^3)$ in the perturbative parameter $\varepsilon$.

Hence,  in this perturbative regime we can work in a manifest SL(2,$\mathbb{Z}$) form by keeping the terms in the action  (\ref{FactionR}) up to order $\calo(\varepsilon^2)$. Up to this order, the presence of the non-trivial warping $e^{-4A}=\varepsilon^{-2}+e^{-4\hat A}$ can be made innocuous by rescaling the fields as follows
\be
\tilde\lambda \rightarrow (1-\frac14\ e^{-4\hat A}\varepsilon^2)\tilde\lambda\quad~~~~ 
\tilde\psi \rightarrow (1+\frac14\ e^{-4\hat A}\varepsilon^2)\tilde\psi\quad~~~~ 
\tilde\rho \rightarrow (1-\frac14\ e^{-4\hat A}\varepsilon^2)\tilde\rho\quad~~~~ 
\ee 
By taking this into account the action (\ref{FactionR}) can be written as
 \be
 \frac12\int_D\d^4\sigma\sqrt{\det h}\,\tilde\Psi \tilde\calt_{\rm F}\tilde\Psi
 \ee 
 where
 \be
  \tilde\Psi\equiv\left(\begin{array}{c} \tilde\lambda \\ \tilde\psi \\ \tilde\rho\end{array}\right)\in S_+\otimes\Big[
  (\Lambda_D^{0,0}\otimes L^{-1}_Q)\oplus   (\Lambda_D^{1,0}\otimes L_Q)\oplus (\Lambda_D^{0,2}\otimes L^{-1}_Q)
  \Big]
 \ee
 and $\tilde\calt_{\rm F}$ is the symmetric operator
 \be
 \begin{aligned}
\calt_{\rm F}&= \calt_{(0)\rm F} +\varepsilon \calt_{(1)\rm F}  +\varepsilon^2 \calt_{(2)\rm F}+\ldots \\
&\equiv 2\left(\begin{array}{ccc} 
0 & \del_Q^\dagger & 0 \\
\delbar_Q & 0 & -\delbar^\dagger \\
0 & -\del & 0
\end{array}\right)+\varepsilon\left(\begin{array}{ccc} 
0 & 0 & 0 \\
0 & \tilde\calr_{(0)} & 0 \\
0 & 0 & \tilde\cals_{(0)}
\end{array}\right)+\varepsilon^2\left(\begin{array}{ccc} 
0 & 0 & 0 \\
0 & 0 & \tilde\calu \\
0 & \tilde\calu^T & 0
\end{array}\right)+\ldots
\end{aligned}
\ee
Here  $\tilde\calu$ is the operator defined in (\ref{calu}), its transposed $\tilde\calu^T:\Lambda^{1,0}_D\rightarrow \Lambda^{2,0}_D$ is just given by $(\tilde\calu_a\wedge \d s^a)\wedge$ and  we have used the leading contributions to the operators $\tilde\cals=\tilde\cals_{(0)}+\varepsilon^2\tilde\cals_{(2)}+\ldots$ and  $\tilde\calr=\tilde\calr_{(0)}+\varepsilon^2\tilde\calr_{(2)}+\ldots$, explicitly given by
\begin{subequations}
\begin{align}
\tilde\cals_{(0)ab}{}^{\bar c\bar d} &= -e^{-\phi}\, \calk^{\bf v}{}_{e[a}\calf_{b]}{}^e\,\Omega_{{\bf v}}{}^{\bar c\bar d}\label{tildes0}\\
\tilde\calr_{(0)\bar a}{}^b&=\,\bar\Omega^{{\bf v}c}{}_{\bar a}\Big[  
e^{-\phi}\,{\bf v}(\phi)\, \calf_c{}^b +\frac{\ii}{2}(G_{2,1})_{{\bf v}c}{}^b \Big]\label{Rexp1}
\end{align}
\end{subequations}

We can now address the problem of solving the zero mode equation
\be
\tilde\calt_{\rm F}\tilde\Psi^{\rm z.m.}=0
\ee
by looking for a perturbative solution
\be
\tilde\Psi^{\rm z.m.}=\tilde\Psi^{\rm z.m.}_{(0)}+\varepsilon\,\tilde\Psi^{\rm z.m.}_{(1)}+\varepsilon^2\,\tilde\Psi^{\rm z.m.}_{(2)}+\ldots
\ee
At first order one gets
\begin{subequations}\label{expzero}
\begin{align}
\tilde\calt_{\rm F(0)}\tilde\Psi^{\rm z.m.}_{(0)}&=0\label{expzero0}\\
\tilde\calt_{\rm F(0)}\tilde\Psi^{\rm z.m.}_{(1)}&=-\tilde\calt_{\rm F(1)}\tilde\Psi^{\rm z.m.}_{(0)}\label{expzero1}
\end{align}
\end{subequations}
The zeroth-order condition (\ref{expzero0}) explicitly reads
\be
\begin{aligned}
\del_Q^\dagger\tilde\psi^{\rm z.m.}_{(0)}&=0\\
\delbar_Q\tilde\lambda^{\rm z.m.}_{(0)}-\delbar_Q^\dagger\tilde\rho^{\rm z.m.}_{(0)}&=0\\
\del_Q\tilde\psi_{(0)}^{\rm z.m.}&=0
\end{aligned}
\ee
Hence, at zeroth order the fluxes have no effect and the spectrum is given by harmonic forms computed by using the U(1)$_Q$-twisted Laplacians $\Delta_{\del_Q}\equiv\del_Q\del_Q^\dagger+\del_Q^\dagger\del_Q$ and $\Delta_{\delbar_Q}\equiv\delbar_Q\delbar^\dagger_Q+\delbar_Q^\dagger\delbar_Q$: 
\be
\tilde\Psi^{\rm z.m.}_{(0)}\in \text{Harm}_{\delbar_Q}^{0,0}(D)\oplus \text{Harm}_{\del_Q}^{1,0}(D)\oplus \text{Harm}_{\delbar_Q}^{0,2}(D)
\ee

At the next order  we need to consider (\ref{expzero1}), which more explicitly reads
\be\label{rhzeros}
\begin{aligned}
\del^\dagger_Q\tilde\psi^{\rm z.m.}_{(1)}&=0\\
\delbar_Q\tilde\lambda^{\rm z.m.}_{(1)}-\delbar^\dagger\tilde\rho^{\rm z.m.}_{(1)}&=\frac12\,\tilde\calr_{(0)}\cdot\psi^{\rm z.m.}_{(0)}\\
\del_Q\tilde\psi^{\rm z.m.}_{(1)}&=\frac12\tilde\cals_{(0)}\cdot\rho^{\rm z.m.}_{(0)}
\end{aligned}
\ee
The condition for the solution of the first and third lines works as in the  discussion on the left-handed sector. Hence a mode $\rho^{\rm z.m.}_{(0)}\in \text{Harm}_{\delbar_Q}^{0,2}(D)$ remains a zero mode at first order only if $\tilde\cals_{(0)}\cdot\rho^{\rm z.m.}_{(0)}$ is $\del_Q$-exact. Otherwise, the zero mode corresponding to $\rho^{\rm z.m.}_{(0)}$ is lifted at first order.

On the other hand, by the Hodge decomposition theorem, we can decompose $\tilde\calr_{(1)}\cdot\psi^{\rm z.m.}_{(0)}$ in a $\delbar_Q$- harmonic piece, a $\delbar_Q$-exact piece and a $\delbar^\dagger_Q$-exact piece.  Hence the second condition in (\ref{rhzeros}) admits a solution only if the $\delbar_Q$- harmonic  component of $\tilde\calr_{(0)}\cdot\tilde\psi^{\rm z.m.}_{(0)}$ is vanishing. Otherwise, also the zero mode corresponding to $\psi^{\rm z.m.}_{(0)}$ is lifted at first order.

We denote by $\calh_{\del_Q}$ and $\calh_{\delbar_Q}$ the projectors on the space of harmonic forms of $\del_Q$ or $\delbar_Q$ respectively and introduce the following linear operators acting on harmonic forms:
\be\label{calptilde}
\begin{aligned}
\tilde\calp_{1,0}\equiv&\ \calh_{\delbar_Q}\circ\tilde\calr_{(0)} :\ \text{Harm}_{\del_Q}^{1,0}(D)\rightarrow \text{Harm}_{\delbar_Q}^{0,1}(D)\\
\tilde\calp_{0,2}\equiv&\ \calh_{\del_Q}\circ\tilde\cals_{(0)}\ :\ \text{Harm}_{\delbar_Q}^{0,2}(D)\rightarrow \text{Harm}_{\del_Q}^{2,0}(D)
\end{aligned}
\ee
We can  summarize the spectrum of the right-handed zero modes at first order in $\varepsilon$ as follows
\be\label{rhzm}
\tilde\Psi^{\rm z.m.}\in \text{Harm}_{\delbar_Q}^{0,0}(D)\oplus \ \ker\tilde\calp_{1,0}\ \oplus\ \ker\tilde\calp_{0,2}
\ee
We see that only  the zero modes corresponding to $\tilde\psi^{\rm z.m.}_{(0)}$ and $\tilde\rho^{\rm z.m.}_{(0)}$ can be lifted 
 by bulk and world-volume fluxes at first order in $\varepsilon$. 

At this point, one could proceed with the second order in $\varepsilon$. At this order, the operator $\tilde\calt_{\rm F(2)}$ mixes $\tilde\psi$ and $\tilde\rho$ and can provide an additional lifting mechanism for the associated zero modes. 
However, we will not pursue the explicit calculation in this paper, since the first order effects can already be the relevant ones for physical applications. In specific cases in which this is not the case, one can proceed with the above perturbative analysis  quite straightforwardly.

In fact, there is an important conclusion which can be reached without much effort. 
The complete action (\ref{FactionR}) does not contain massive terms involving $\tilde\lambda$. Hence, if we used (\ref{FactionR}) for  studying the zero modes  $\tilde\lambda^{\rm z.m.}$, we would conclude that (at any order in $\varepsilon$) fluxes have no effect on them and they would still be counted by  $ \text{Harm}_{\delbar_Q}^{0,0}(D)$.
We are implicitly assuming that the subleading problems with SL(2,$\mathbb{Z}$) duality, being relevant just for the $\tilde\psi$ zero modes,  cannot spoil this conclusion.  In support to this assumption, notice that the SL(2,$\mathbb{Z}$) non-invariance does not constitute a practical problem in the orientifold limit. In this limit the zero modes  $\tilde\lambda^{\rm z.m.}$ coincide with pseudo-universal zero modes often denoted by $\bar\tau^{\dot\alpha}$ -- see \cite{Bianchi:2011qh} for a more detailed discussion. In this limit we can more comfortably state that  fluxes cannot lift  the $\tilde\lambda^{\rm z.m.}\sim\bar\tau^{\dot\alpha}$ zero modes, at least  within the supergravity approach we are adopting.

\subsection{Zero modes and supersymmetry breaking}
\label{sec:SBmodes}

Let us now consider what happens to the zero modes if we include a supersymmetry breaking flux component $G_{0,3}\neq 0$.
This is easily seen from the contribution $\Delta^{\rm{SB}}S^{\rm ferm}$ given in (\ref{SBaction}). 

By rescaling $\lambda$ as in (\ref{warprescal}) the warping disappears from the left-moving contribution $\Delta^{\rm SB}S^{\rm ferm}_{\rm L}$. Then, we see that the $G_{0,3}\neq 0$ flux just provides a mass term proportional to $\bar{G_{\it 3}}\cdot\bar\Omega$ for $\lambda$ and, as expected, the breaking of supersymmetry lifts the universal zero modes $\lambda^{\rm z.m.}\sim \theta^{\alpha}$. One may suspect that, being the bulk supersymmetry broken, the effective interactions induced by the instanton cannot manifestly  be described  as F-terms. We will come back to this point when we will discuss the effective four-dimensional superpotential in section \ref{sec:supgen}.

In the right-handed sector we see a similar effect: $G_{0,3}\neq 0$ induces a mass term for $\tilde\lambda$, which would mean that $G_{0,3}\neq 0$ lifts the $\tilde\lambda^{\rm z.m.}\sim\bar\tau^{\dot\alpha}$ zero modes.  Usually, the presence of the pseudo-universal zero modes $\bar\tau^{\dot\alpha}$ is problematic for several applications,  as for instance it prevents the generation of a superpotential by the instanton. Hence, the coupling $\Delta^{\rm{SB}}S^{\rm ferm}_{\rm R}$ appears to be helpful to alleviate this problem. On the other hand, one has to keep in mind that such a term is actually generated by the supersymmetry-breaking flux. Furthermore,  in the large  volume expansion of section (\ref{sec:pert}) this term is of order $\calo(\varepsilon^3)$ and so is much suppressed with respect to the term which lifts the universal zero modes $\lambda^{\rm z.m.}\sim \theta^{\alpha}$ discussed above. So, it is not at all obvious that such a mechanism can be helpful for generating desired interactions of an effective supersymmetric theory. 

Nevertheless, one could consider a regime in which this term is not subleading. To understand this point, suppose for simplicity that the restriction of the bundle $L_Q$ to $D$ is trivial. This means that $\tau$ is constant on $D$ and $Q_{\it 1}|_D=0$, so that $\tilde\lambda^{\rm z.m.}\sim\bar\tau^{\dot\alpha}$ are just constant modes on $D$. Then, by plugging $\lambda=\theta$ and $\tilde\lambda=\bar\tau$ in (\ref{SBaction}), we get the following schematic effective mass terms for $\theta$ and $\bar\tau$, in the large breathing mode limit:
\be\label{effSB}
S^{\rm SB}_{\rm eff}\sim \varepsilon\,\theta\theta\,\text{Vol}(D)\,\bar{G_{\it 3}}\cdot\bar\Omega +\varepsilon^3\,\bar\tau\bar\tau\,G_{\it 3}\cdot\Omega\,\int_D\calf\wedge \calf\,
\ee
where $\text{Vol}(D)$ is measured by the induced K\"ahler metric.
We could now suppose that the volume  $\text{Vol}(D)$ can be taken to be very small. The $\bar\tau$ mass in term in (\ref{effSB}) does not depend on $\text{Vol}(D)$  and, if for instance $\calf$ is a pure non-trivial U(1) field-strength,  $\int_D\calf\wedge \calf$ is a topological quantity which does not change as the cycle shrinks. 

Hence, in such a regime the second term in (\ref{effSB}) could dominate on the first one. 
Even though the supergravity approximation is not in principle valid in this regime, it is reassuring that our result is perfectly consistent with the 
calculations on fractional D(-1)-instantons at orbifold singularities performed in \cite{Billo:2008sp}. Indeed, if $D$ is the exceptional divisor providing the blow-up of an orbifold singularity,  in  the strict $\text{Vol}(D)\rightarrow 0$ limit the E3-brane becomes a fractional D(-1)-instanton at the orbifold singularity and the effective action (\ref{effSB}) reduces just to a term of the form $(G_{\it 3}\cdot\Omega)\bar\tau\bar\tau$. Indeed, the authors of \cite{Billo:2008sp} found such a term and no mass-term for $\theta$, in agreement with our result.

\subsection{Cohomological summary}

Let us restate our results in terms of cohomology classes. The fermionic zero modes can be identified with (twisted) harmonic forms which obey some additional projection conditions. We can use the following isomorphism between  harmonic forms  and cohomology classes 
\be
\begin{array}{l c l c l}
\text{Harm}^{0,p}_{\delbar}(D)&\simeq& H^{0,p}_{\delbar}(D) &\simeq& H^p(D,\calo_D)\cr
\text{Harm}^{p,0}_{\del}(D)&\simeq& H^{p,0}_{\del}(D)&\simeq& H^p(D,\bar \calo_D)\cr
\text{Harm}^{0,p}_{\delbar_Q}(D)&\simeq& H^{0,p}_{\delbar_Q}(D) &\simeq& H^p(D,\call_Q^{-1})\cr
\text{Harm}^{p,0}_{\del_Q}(D)&\simeq& H^{p,0}_{\del_Q}(D) &\simeq& H^p(D,\bar\call_Q^{-1})
\end{array}
\ee
The third entry in each line contains \v{C}ech cohomology groups  associated with the trivial line bundle $\calo_D$ on $D$ and the holomorphic line bundle $\call_Q^{-1}\simeq K_X$ (restricted to $D$), as well as their complex conjugate anti-holomorphic bundle $\bar\calo_D$ and $\bar\call_Q^{-1}$. By using this isomorphism we can take the operators $\calp_{2,0}$, $\tilde\calp_{1,0}$ and $\tilde\calp_{0,2}$, defined in (\ref{calp}) and (\ref{calptilde}), as acting on the above cohomology groups.

Hence, for supersymmetric bulks ($G_{0,3}=0$), the spectrum of fermionic zero modes can be summarized as follows: 
\be
\begin{aligned}
\label{cohozm}
&\begin{array}{|c|c|} \text{l.h.\ fermions}& \text{zero modes}\\ \hline
\lambda^\alpha&  H^0(D,\bar\calo_D)  \\
\psi^\alpha &   H^1(D,\calo_D)\\
\rho^\alpha &  \ker \calp_{2,0} \\
\end{array}
\quad~~~
\begin{array}{|c|c|} \text{r.h.\ fermions}& \text{zero modes}\\ \hline
\tilde\lambda^{\dot\alpha}&  H^0(D,\call^{-1}_Q)  \\
\tilde\psi^{\dot\alpha} &   \ker\tilde\calp_{1,0} \\
\tilde\rho^{\dot\alpha} &  \ker \tilde\calp_{0,2}
\end{array}
\end{aligned}
\ee
where $\ker \calp_{2,0}\subset H^2(D,\bar\calo_D)$, $\ker\tilde\calp_{1,0}\subset H^1(D,\bar\call^{-1}_Q) $ and $\ker \tilde\calp_{0,2}\subset H^2(D,\call_Q^{-1}) $.

Notice that all quantities have a counterpart in the orientifold limit, in which one works the double cover divisor $\tilde D$ in the double-cover CY three-fold $\tilde X$, see section \ref{sec:Oplanes}. Indeed, all the results are readily adapted to this limiting case by using the identifications $H^i(D,\calo_D)\simeq H^{0,i}_+(\tilde D)$ and $H^i(D,\call_Q^{-1})\simeq H^{0,i}_-(\tilde D)$, where we have used the split $H^{p,q}(\tilde D)=H^{p,q}_+(\tilde D)\oplus H^{p,q}_-(\tilde D)$ of $(p,q)$-cohomology groups into subgroups  whose elements are even/odd under the orientifold involution. 

Notice that by setting $\tau$ constant, with or without orientifolds, and forcing $\calf=0$ one recovers, as a particular subcase, the zero-mode structure found in \cite{Bergshoeff:2005yp} (up to a change of conventions). 

By turning on a supersymmetry breaking flux $G_{0,3}\neq 0$ the universal zero modes $(\lambda^{\rm z.m.})^\alpha\sim \theta^\alpha$ are lifted, as expected from their role as goldstini. The pseudo-universal zero modes $(\tilde\lambda^{\rm z.m.})^{\dot\alpha}\sim \bar\tau^{\dot\alpha}$ are lifted too, but at subleading $\calo(\varepsilon^3)$ order in the large universal modulus expansion of section \ref{sec:pert}. On the other hand this $\bar\tau^{\dot\alpha}$-lifting effect could become more relevant as the volume of the divisor becomes  smaller.

\section{Geometric interpretation of fermionic zero modes}
\label{sec:geomint}

Part of the flux induced terms discussed in the previous section, which can lift zero modes, admit a geometric interpretation. This is expected by supersymmetry, which relates fermionic zero modes to geometrical bosonic ones.  In the absence of bulk fluxes, this interpretation has been already discussed in detail in section 4.6  of \cite{Bianchi:2011qh} -- see also appendix B therein for a summary of useful results on deformations of holomorphic cycles and extrinsic curvature.

\subsection{Geometric zero mode lifting}

Let us work in the large universal modulus regime, in which the warping is negligible. As we have discussed, this limit does not change at all
the structure of the left moving sector, while it simplifies the computation of the right moving spectrum, still preserving a substantial part of its non-trivial features.   

First, let us consider a zero mode $\tilde\rho^{\rm z.m.}$. Recall that $\tilde\rho^{\rm z.m.}\in \text{Harm}^{0,2}_{\delbar_Q}(D)$ and in addition $[\tilde\cals_{(0)}\cdot \tilde\rho^{\rm z.m.}]=0$ in $\del_Q$-cohomology, with $\tilde\cals_{(0)}$ as defined in (\ref{tildes0}).
One can associate with $\rho^{\rm z.m.}$ a normal (0,1)  vector $ \tilde V\in N^{0,1}_D$ defined by
\be
\tilde V^{\bar{\bf v}}=\frac12\, e^{-\frac\phi2}\Omega^{\bar{{\bf v}}\bar a\bar b}\,\tilde\rho^{\rm z.m}_{\bar a\bar b}
\ee
Since $\Omega$ is U(1)$_Q$ covariantly constant and $\tilde\rho^{\rm z.m.}$ is harmonic, then $\tilde V\in H^{0}_{\del}(D, N^{0,1}_D)$.
$\tilde V$ can be seen as the fermionic supersymmetric partner of a generator of an anti-holomorphic deformation of the E3-brane embedding. Such a deformation can induce a deformation of the complex structure on $D$  inherited  by the bulk, which can in turn generate a supersymmetry-violating $(2,0)$ component of $\calf$. One can then borrow the discussion of section 4.6 of \cite{Bianchi:2011qh} to conclude that such deformation is  given by
\be
\delta_{\tilde V}^{\rm c.s.}(e^{-\frac\phi2}\calf)^{2,0}=-\,e^{-\frac{\phi}{2}} \calk^{{\bf v}}{}_{ e[ a}\calf_{ b]}{}^{ e}\, V_{{\bf v}}\d s^a\wedge \d s^b\equiv 2\tilde\cals_{(0)}\cdot\tilde\rho^{\rm z.m.}
\ee
This suggests an immediate interpretation of the cohomological condition $[\tilde\cals\cdot \tilde\rho^{\rm z.m.}]=0$.
Indeed, it means that the (2,0) component of  the deformed $e^{-\frac\phi2}\calf$ can be reabsorbed by a 
$\delbar_Q$-exact piece, which would be interpretable as a deformation of the world-volume gauge field which does not change the topological nature of the flux. This would then neutralize the potential supersymmetry-breaking effect of the geometric deformation, preserving it in the spectrum of zero modes. Such interpretation becomes sharper in the orientifold limit, in which everything can be translated in terms of an ordinary gauge bundle supported in the double-cover divisor $\tilde D$. On the other hand, a non-exact $\tilde\cals\cdot~\!\!\!\tilde\rho^{\rm z.m.}$ would correspond to a 
topologically non-trivial $\delta_{\tilde V}^{\rm c.s.}(e^{-\frac\phi2}\calf)^{2,0}$ which could not be reabsorbed by any gauge field deformation and would then break the supersymmetry condition $\calf^{2,0}=0$, leading to the lift of the associated zero mode.

Let us now consider the left-handed sector.
A zero mode $\rho^{\rm z.m.}$ is defined by
$\rho^{\rm z.m.}\in \text{Harm}^{2,0}_{\del}(D)$ and $[\cals\cdot \rho^{\rm z.m.}]=0$ in $\delbar$-cohomology.
It is naturally associated with a U(1)$_Q$-twisted normal (1,0) vector $ V\in N^{1,0}_D\otimes L_Q^{-1}$ defined by:
\be
V^{\bf v}=\frac12\, e^{-\frac\phi2}\,\bar\Omega^{{\bf v} a b}\,\rho^{\rm z.m}_{a b}
\ee
In analogy with the above discussion, we can think of $V$ as the fermionic supersymmetric partner of a U(1)$_Q$-twisted holomorphic deformation of the holomorphic embedding of the E3-brane. In the orientifold limit, this reduces to a deformation of the double cover divisor $\tilde D$ which is {\em odd} under the orientifold involution. Such a deformation would induce a possible supersymmetry-violating (0,2) component of $\calf$:
\be
\delta_{V}^{\rm c.s.}(e^{-\frac\phi2}\calf)^{0,2}=-\,e^{-\frac{\phi}{2}} \calk^{\bar{\bf v}}{}_{\bar e[\bar a}\calf_{\bar b]}{}^{\bar e}\, V_{\bar{\bf v}}\,\d\bar s^{\bar a}\wedge \d\bar s^{\bar b}
\ee
On the other hand, one has to be careful about  the possible contribution of the bulk-fluxes to $\delta_V\calf^{0,2}$ . 
Such a contribution has been purposely omitted in the above discussion about  $\tilde\rho^{\rm z.m.}$
since, as we will see, in that case it turns out to be vanishing. 

In order to preserve the Bianchi identity $\d\calf=H_{\it 3}|_D$, an embedding  deformation generated by $V$
must be compensated by a deformation $\delta^{\rm emb}_V\calf=\iota_V H_{\it 3}$. On the other hand,
the dual field-strength $\calf_{\rm D}$ given in section \ref{sec:E3bos}  should satisfy the dual Bianchi identity $\d\calf_{\rm D}=\tilde F_{\it 3}|_D$ (since $\tilde F_{\it 3}=\d C_{\it 2}$ locally).  Hence, analogously, $\delta^{\rm emb}_V\calf_{\rm D}=\iota_V \tilde F_{\it 3}$. We could now use these observations to compute the corresponding deformation $\delta^{\rm emb}_V\calf_+$ of the self-dual flux $\calf_+$ as defined in (\ref{sdfluxes}). By applying the result to a supersymmetric flux, for which $\calf_+\equiv \calf$, we find
\be
\delta^{\rm emb}_V(e^{-\frac{\phi}{2}}\calf)^{0,2}=\frac{\ii}{2}e^{\frac{\phi}{2}}\iota_V\bar{G_{2,1}}
\ee
Hence, in total one gets
\be
\delta_V(e^{-\frac{\phi}{2}}\calf)^{0,2}=(\delta^{\rm c.s}_V+\delta^{\rm emb}_V)(e^{-\frac{\phi}{2}}\calf)^{0,2}\equiv 2\cals\cdot\rho^{\rm z.m.}
\ee
with $\cals$ given in (\ref{S1}). On the other hand, one could analogously compute $\delta^{\rm emb}_{\tilde V}(e^{-\frac{\phi}{2}}\calf)^{2,0}=\frac{\ii}{2}e^{\frac{\phi}{2}}\iota_{\tilde V}\bar{G_{2,1}}\equiv 0$, confirming the validity of the above discussion about $\tilde\rho^{\rm z.m.}$.

One can now proceed by interpreting the zero mode lifting condition for $[\cals\cdot \rho^{\rm z.m.}]=0$ as we did for $\tilde\rho^{\rm z.m.}$.
Namely, only the geometric deformations $V$ which generate  a cohomologically non-trivial  (2,0) component of $e^{-\frac{\phi}{2}}\calf$ correspond to lifted geometric modes.\footnote{These geometrical ways of interpreting the flux-induced zero mode lifting  could be regarded as the F-theory E3-brane counterpart of what found in \cite{Gomis:2005wc,luca2,Koerber:2006hh} about the deformations of space-filling D7-branes in flux compactifications.}  

At order $\calo(\varepsilon)$, there is also another mass term of the form $\tilde\psi\cdot\tilde\calr_{(0)}\cdot\tilde\psi$, with $\tilde\calr_{(0)}$
given in (\ref{Rexp1}). Unfortunately, a geometrical interpretation of the associated  zero modes lifting is less clear.

\subsection{When is there no flux-induced zero mode lifting?}

In some cases, the fluxes turn out not to affect the fate of certain zero modes.
As a preliminary remark, let us first notice that the Bianchi identity $\d \calf=H_{\it 3}|_D$ and its dual $\d\calf_{\rm D}=\tilde F_{\it 3}|_D$ imply the following equations for the supersymmetric flux
\begin{subequations}\label{covBI}
\begin{align}
\delbar_Q(e^{-\frac{\phi}{2}}\calf)&=\frac{\ii}{2}e^{\frac{\phi}{2}}\bar{G_{2,1}}|_D\label{covBI1}\\
\del_Q(e^{-\frac{\phi}{2}}\calf)&=0\label{covBI2}
\end{align}
\end{subequations}
which can be more clearly obtained by first writing the  equations for general $\calf_+$ and $\calf_-$ and then setting $\calf_-=0$ and $\calf_+=\calf$.%
\footnote{\label{foot:BI}This clarifies some confusion in \cite{Bianchi:2011qh}, in which the  condition (\ref{covBI2}) was argued not to arise from $\d\calf=H_{\it 3}|_D$ alone but to be the most natural one.  In addition to (\ref{covBI}), there  is another equation which is implied by the Bianchi and dual Bianchi identities: $\del\tau|_D\wedge \calf+G_{2,1}|_D=0$. This is just the gauge field equation of motion (\ref{Aeom})  which is not automatically guaranteed by the supersymmetry conditions. So, only the combinations (\ref{covBI}) of Bianchi's and the dual Bianchi's are really automatically satisfied.}

There are at least  two cases in which there is no flux-induced lifting for the $\rho$ and $\tilde\rho$ zero modes. 
First of all, notice that, by definition, 
\be
\begin{aligned}
-\, V_{\bar{\bf v}}\,\calk^{\bar{\bf v}  a}{}_{\bar b}\, \frac{\del}{\del s^{ a}}\otimes\d \bar s^{\bar b}&\equiv (\bar \nabla_Q V)_\|\\
-\,\tilde V_{{\bf v}}\,\calk^{{\bf v} \bar a}{}_{b}\, \frac{\del}{\del \bar s^{\bar a}}\otimes\d s^{b}&\equiv (\nabla \tilde V)_\|
\end{aligned}
\ee
where $\nabla_Q $ ($\bar\nabla_Q$) is the bulk (anti-)holomorphic covariant differential and by ${}_\|$ we mean that we are projecting the indices along $D$. 

A first case in which there is no flux-induced zero mode lifting occurs when $e^{-\frac{\phi}{2}}\calf$ can be extended
to a $(1,1)$ form $e^{-\frac{\phi}{2}}\hat\calf$, of U(1)$_Q$-charge $q_Q=-1$, which lives on a neighbourhood of $D$ and is $\del_Q$-closed.  Indeed, in this case we can write
\be
\begin{aligned}
\tilde\cals_{(0)}\cdot \tilde\rho^{\rm z.m.}&=-\frac12e^{-\frac{\phi}{2}}\, \tilde V_{{\bf v}}\,\calk^{{\bf v}\bar c}{}_{a}\calf_{b \bar c}\, \d  s^{a}\wedge \d  s^{b}\\
&\equiv-\frac12e^{-\frac{\phi}{2}}\,(\nabla \tilde V)^{\bar c}{}_a\,  \calf_{\bar c b}\, \d  s^{a}\wedge \d  s^{b}=- \frac12e^{-\frac{\phi}{2}}\,(\nabla \tilde V)^{\bar m}{}_a\, \hat\calf_{\bar m b}\, \d  s^{a}\wedge \d  s^{b}\\
&=-\frac12\del_Q(e^{-\frac{\phi}{2}}\iota_{\tilde V}\hat \calf)|_D
\end{aligned}
\ee
Hence, we see that indeed $\tilde\cals_{(0)}\cdot \tilde\rho^{\rm z.m.}$ is trivial in $\del_Q$-cohomology and then the zero mode $\tilde\rho^{\rm z.m.}\in \text{Harm}^{0,2}_{\delbar_Q}(D)$ is not lifted by fluxes.

By the very same reasoning,  a zero mode $\rho^{\rm z.m.}\in \text{Harm}^{2,0}_{\del}(D)$  is not lifted by fluxes if  the extension $e^{-\frac{\phi}{2}}\hat\calf$ satisfies the condition $\delbar_Q(e^{-\frac{\phi}{2}}\calf)=\frac{\ii}{2}e^{\frac{\phi}{2}}\bar{G_{2,1}}$, that is if it fulfills the extension of  (\ref{covBI1}) outside $D$.

These two results can be combined into a stronger statement: if the world-volume flux can be extended to a neighbourhood of  $D$ while still preserving the (extended) Bianchi identities (\ref{covBI}) then the bulk and world-volume fluxes have no effect on the   $\rho^{\rm z.m.}$ and $\tilde\rho^{\rm z.m.}$.

There is another interesting subcase in which the effect of fluxes becomes weaker. Suppose that we can uplift the anti-holomoprhic section $\tilde V$ of $N^{0,1}_D$ to an anti-holomorphic section of $T^{0,1}_X|_D$.  Then, we could write $\nabla \tilde V=\nabla \tilde V^{\bar\imath }\,\frac{\del}{\del \bar z^{\bar\imath}}=\del_a\tilde V^{\bar\imath }\,\d s^{a}\otimes \frac{\del}{\del\bar z^{\bar \imath}}=0$ and then $\tilde S_{(0)}\cdot\tilde\rho^{\rm z.m.}$ would identically vanish.  For instance, this happens if  we can {\em holomorphically} split $T^{1,0}_X|_D\simeq_{\rm hol}T_D^{1,0}\otimes N^{1,0}_D$, as it is indeed the case when $N^{1,0}_D$ is trivial. On the other hand, such topological assumptions are not sufficient to conclude that there is no lift of the $\rho^{\rm z.m.}$ modes, since in that case the operator $\cals$ still contains the bulk flux, whose effect is not necessarily trivial.

On the other hand, there does not seem to be an analogous geometrical/topological condition  which would prevent the lift of the $\tilde\psi$ zero modes. 

\section{Applications to non-perturbative F-terms}
\label{sec:supgen}

In the above sections we have focused on the calculation of fermionic zero modes on supersymmetric E3-brane instantons.
One of the main possible applications is, of course, the computation of non-perturbative contributions to the effective four-dimensional superpotential or of more general F-terms. The logic behind this approach has been explained in  \cite{Becker:1995kb,Harvey:1999as} and here we would like to just sketch a few possible applications of our results, leaving more explicit computations to the future.  As in the rest of the paper, we will mostly restrict to the case of supersymmetric vacua, postponing the discussion of the case $G_{0,3}\neq 0$ to section \ref{sec:SB}.

The dimensional reduction of the ten-dimensional string theory produces a four-dimensional effective action which is necessarily supersymmetric. The complex structure and axio-dilaton moduli are generically lifted at tree-level by the bulk fluxes and then do not appear as
low-energy  chiral fields in the four-dimension effective Lagrangian. 

On the other hand, in the low-energy spectrum there are other chiral fields, let us call them $T^I$, which cannot appear in the low-energy effective superpotential. The reason is that, at the perturbative level, string theory is  invariant under constant shifts $T^I\rightarrow T^I+\ii\,c^I$, $c^I\in\mathbb{R}$. Hence, by holomorphy, the four-dimensional perturbative superpotential cannot depend on $T^I$. On the other hand, at the non-perturbative level the continuos symmetry is actually broken to $c^I\in \mathbb{Z}$, and then the superpotential can depend on combinations of $e^{-2\pi\,n_I T^I}$, with $n_I\in \mathbb{Z}$. 

Now, the contribution of an E3-brane instanton to the path integral  can be formally written as 
\be\label{branePI}
\int\cald X\cald\Theta\,  e^{- S_{\rm E3}[\Phi(X,\Theta)]}
\ee
where $X$ symbolically denotes all world-volume bosonic fields, while $\Theta$ denotes the world-volume fermions. 
Clearly, the integration over fermions plays a crucial role in establishing which are the potential contributions. 
We only consider the one-loop approximation in which only quadratic fluctuations around the instantonic configuration are taken into account. Higher order terms in the fermions may lift some of the one-loop zero-modes but we will not study them here.

On the other hand, there are also fermionic bilinear terms which couple  world-volume and bulk fermions.  The corresponding contribution to the E3 effective action can be found in section 9  of \cite{Lust:2008zd} and, by using the $\kappa$-fixing (\ref{kappafixing}) for the GS-fermions $\Theta$,  reads
\be\label{intferm}
S_{\rm int}=\ii\int_D\d^4\sigma\, e^{-\phi}\sqrt{\det \calm}
\, \Theta^T\calc\Big[(\calm_\sigma^{-1})^{AB}\Gamma_A
\Psi_B-\frac12\Xi\Big]
\ee  
where $\Psi_A$ is the pullback   to the world-volume $D$ of the bulk (doubled) type II gravitino $\Psi_M$ and $\Xi$ is the bulk dilatino. 
The interactions encoded in (\ref{intferm}) play a crucial role in determining the  contribution of the instanton to the low-energy effective action. 

\bigskip

{\em Generation of superpotentials}

\bigskip

\noindent In particular, we can see how it works for the superpotential. As we have said, we have restricted 
the bosonic configuration to be the supersymmetric E3-brane sitting at a certain point $x^\mu\in \mathbb{R}^4$. Take the fermionic field $\Theta$ to be just the universal zero mode:
\be\label{Fzeromodesansatz}
\Theta_1=e^{\frac{A}{2}+\frac{\phi}{8}}\theta\otimes\eta^* \, ,\qquad \Theta_2=\ii e^{\frac{A}{2}+\frac{\phi}{8}}\theta\otimes\eta^* 
\ee
where $\theta\equiv \lambda^{\rm z.m.}$ is a chiral spinor $\theta\equiv\theta^\alpha$ which is constant along $D$. (We are implicitly assuming that $D$ is a connected irreducible holomorphic hypersurface, so that dim Harm$^{0,0}_\del(D)=0$.) Then, the instanton GS effective action gives a four-dimensional chiral superfield by construction:
\be\label{expS}
S[\Phi(X,\Theta)]\equiv T(x,\theta)=t(x)+\theta\psi_T(x)+\theta^2\,F_T(x)
\ee
where $\theta\psi_T(x)=\theta^\alpha \psi_{T\alpha}(x)$ and $\theta^2=\theta^\alpha\theta_\alpha$. For E3-branes in a supersymmetric bulk ($G_{0,3}=0$) the tree-level expectation value of the F-term of $T$ vanishes, $ \langle F_T\rangle_{\rm tree}=0$.  
The chiral field $T(x,\theta)$ is just the closed string  low-energy  chiral field (with $\bar\theta^{\dot\alpha}=0$) which couples directly to the supersymmetric E3-brane instanton. On the other hand, we could read the precise form of $\psi_T$ in terms of the ten-dimensional fermions by comparing (\ref{expS}) and (\ref{intferm}) with $\Theta$ as in (\ref{Fzeromodesansatz}). This form for $\psi_T$ is predicted by four-dimensional supersymmetry arguments. It would be interesting to compare it with more standard KK techniques, but we will not pursue this procedure in this paper.

In the path-integral  (\ref{branePI}) one can factorize the integral over the position on $x^\mu$, while
a  pair of interaction terms $\theta\psi_T$  in (\ref{expS})  is pull-down from $e^{-S[\Phi(X,\Theta)]}$ and used to absorb the integration measure $\d^2\theta$ of the universal zero modes. If there are no other fermionic and bosonic zero modes, the path-integral (\ref{branePI}) produces the following contribution
\be\label{ins}
\int \d^4 x\,\psi_T(x)\psi_T(x)\,\cala\, e^{-t(x)}
\ee
where $\cala$ contains the contribution of the non-zero modes. Since the instanton preserves the right-handed bulk supersymmetry, we expect the right-handed sector to be supersymmetric, hence providing no contribution to $\cala$.
Hence, $\cala$ is expected to contain just determinants and pfaffians of the left-handed sector. In particular, the relevant fermionic contribution to the prefactor  $\cala$ should be provided just by the Pfaffian of the hermitian operator $\calt_{\rm F}^\dagger\calt_{\rm F}$, where $\calt_{\rm F}$ is defined in (\ref{calt}). Such a simplification happens for instance in the computation of F-terms induced by world-sheet instantons, see e.g.\  \cite{Witten:1999eg,Beasley:2005iu}.

The term (\ref{ins}) should be regarded as an insertion in the low-energy closed string path-integral and exactly matches the insertion produced
by an effective F-term $\int\d^2\theta\, W_{\rm np}$, with
\be\label{npsup}
W_{\rm np}=\cala\,e^{- T}
\ee 
in the low-energy effective action. In other words, as stressed for instance in \cite{Witten:1999eg}, the contribution $\int\d^4x\d^2\theta\, W_{\rm np}$ to the four-dimensional effective action is just given by the brane instanton partition function.
Notice that we are working around a bulk configuration which has a flat four-dimensional space-time, hence getting expressions which naturally fit into a rigid supersymmetric theory  but which naturally generalize to curved space.

One important point is that in general there are many supersymmetric  E3-brane configurations, wrapping different cycles which in turn can support different world-volume fluxes. In particular, several can contribute to the superpotential, namely the ones which have precisely two zero modes, and each of them gives a different contribution. One of the goals of our paper is to provide explicit and complete formulas to decide which single brane instantons do contribute to the superpotential and which do not (at one-loop level). On top of that, there could be additional contributions to the superpotential coming from multi-instanton effects as well as from the coupling to charged chiral matter supported on space-filling branes, which we do not consider in the present paper. The final effective superpotential is the sum of {\em all} such contributions. 
In particular, one has to sum  over all possible E3-configurations, which are specified by {\em both} the embedding and the  world-volume flux. 
The contribution to the path-integral coming from different E3-branes wrapping  the same divisor $D$ but supporting different  world-volume fluxes has been recently studied in \cite{Grimm:2011dj} and \cite{Kerstan:2012cy} by working in the weakly coupled orientifold limit and in the dual M-theory description respectively. The strategy  is complementary to the one based on holomorphic factorization pursued in \cite{Witten:1996hc} and in \cite{Kerstan:2012cy} it is shown  that the results match well. It would be important to combine these approaches with the results of the present paper in order to understand better the structure of the complete superpotential  in the presence of fluxes.\footnote{It would be also interesting to use the general results of this paper to inspect the effects of E3-brane instanton on the internal geometry and its physical implications, along the lines of \cite{eff1,Koerber:2008sx,Marchesano:2009rz,Baumann:2010sx,Dymarsky:2010mf,Aparicio:2011jx}.}

\bigskip

{\em Generation of multi-fermion terms}

\bigskip

\noindent On the other hand, if in addition to the universal zero modes there are other world-volume zero modes, 
then (\ref{ins}) would be supplemented by an additional overall integration over them. 
Of course, in order to get a non-vanishing result one needs to soak up the path-integration over the additional  fermionic zero-modes.
As a result  of the coupling (\ref{intferm}) between bulk and E3-brane fermions the instanton produces multi-fermion terms of the kind discussed in \cite{Beasley:2005iu}.

As a simple example, such a higher order F-term is generated in the presence of two pseudo-universal zero modes $\bar\tau^{\dot\alpha}$. Recalling (\ref{cohozm}), this happens if for instance the restriction of the bundle $\call^{-1}_Q\simeq K_X$ to $D$ is trivial and in this case there are exactly two additional zero-modes, one for each spinorial Weyl index. In the orientifold limit, this happens in the cases in which the E3-brane is rigid with no Wilson lines and does not intersect the $\Omega$-planes, in other words it is  U(1) rather than O(1). 

By using the explicit decomposition (\ref{Fsplit}) in (\ref{intferm}), one gets an effective interaction of the form $\bar\tau\, \bar\chi$, where $\bar\chi$ is a well defined combination of bulk fermions. By four-dimensional supersymmetry arguments, $\chi$ can be seen as the fermionic component of a four-dimensional chiral field $\Sigma$. Furthermore,  by extending the results of \cite{Blumenhagen:2007bn} -- see also \cite{GarciaEtxebarria:2007zv,GarciaEtxebarria:2008pi} --  to our F-theory setting, $\Sigma$ should be a combination of the chiral fields $T^I$. 

Then, by four-dimensional supersymmetry, the effective interaction $\bar\tau\, \bar\chi$ can be completed into $\bar\tau^{\dot\alpha}\bar D_{\dot\alpha}\bar\Sigma$ and such a term in the path-integral produces an higher order F-term of the schematic form
\be
\int\d^4x\,\d^2\theta\,\d^2\bar\tau\, e^{- T+\bar\tau\bar D\bar\Sigma}=\int\d^4x\,\d^2\theta\, \bar D_{\dot\alpha}\bar\Sigma\,\bar D^{\dot\alpha}\bar\Sigma\, e^{-T}
\ee
By using (\ref{Fsplit}) and (\ref{intferm}), together with an explicit KK-ansatz for the closed string modes, the form of this and more general terms can be made quite explicit, but we refrain from doing it in the present paper. 
See e.g.\ \cite{Beasley:2005iu,Blumenhagen:2007bn,GarciaEtxebarria:2007zv,GarciaEtxebarria:2008pi,Uranga:2008nh}, for further discussions on multi-fermion F-terms generated by brane instantons.

\bigskip

{\em Perturbative evaluation of the effect of fluxes}

\bigskip

\noindent  
The computation of F-terms sketched above offers an alternative more physical viewpoint on the perturbative expansion described in section \ref{sec:pert}.  In the large universal modulus limit all the combinations of chiral fields which appear in the exponent $e^{-T}$ of the F-term have very large real parts, $\Re T\sim \varepsilon^{-2}\gg 1$. This is  exactly the region of the moduli space in which we expect the semi-classical approach we are adopting  to be fully trustable.  On the other hand, by the usual holomorphy arguments, the perturbative contribution to the prefactor $\cala$ appearing in the F-term cannot depend on the axionic-like K\"ahler moduli $T^I$. 
Hence, four-dimensional supersymmetry arguments predict that if $\cala$ is non-vanishing at a certain order in $\varepsilon$, it is non-vanishing at all orders and its value does not depend on $\varepsilon$ itself.  In particular, we can perform the $\varepsilon$-perturbative expansion already at the path-integral level and evaluate the result perturbatively. 

In order to decide whether the fluxes can help in generating a superpotential, one can first consider the fermionic  zero-modes at zeroth-order in $\varepsilon$, which are just those corresponding to an unmagnetized E3-brane on a flux-less F-theory compactification. These are just given by  harmonic representatives of $\delbar$ and $\delbar_Q$ cohomology classes or their complex conjugated.
By collectively denoting by $\mu_l$ and $\tilde\mu_r$ these harmonic modes, at the lowest order in $\calo(\varepsilon)$  the relevant contribution to the path-integral is just 
\be
\int \prod_{l,r}\d\mu_l\d\tilde\mu_r \, e^{-\varepsilon\, \sum_{l,m}(\calt^{(1)}_{{\rm F}})_{lm}\mu_l\mu_m-\varepsilon\, \sum_{r,s}(\tilde\calt^{(1)}_{\rm F})_{rs}\tilde\mu_r\tilde\mu_s}\sim \varepsilon^2(\det\calt^{(1)}_{\rm F})_{\rm harm}(\det\tilde\calt^{(1)}_{\rm F})_{\rm harm}
\ee
Here $\calt^{(1)}_{lm}$ and $\tilde\calt^{(1)}_{rs}$  are the matrices corresponding to the restriction to the harmonic sector of the first $\varepsilon$-order operators obtained by expanding in $\varepsilon$  the operators $\calt_{\rm F}$ and $\tilde\calt_{\rm F}$ introduced in section \ref{sec:zm}. 

In particular, the number of residual zero-modes is given by the dimension of the kernels of  the matrices $\calt^{(1)}_{lm}$ and $\tilde\calt^{(1)}_{rs}$. One can easily check that this precisely agrees with the zero mode spectrum  (\ref{lhzm}) and (\ref{rhzm}), which were obtained by direct inspection of the complete fermionic equations of motion in section \ref{sec:zm}.  In particular, the left-moving spectrum is completely fixed at order $\calo(\varepsilon)$ and cannot be changed by higher-order terms in the quadratic fermionic action.
 
Let us remark once again that, if there are $\tilde\lambda^{\rm z.m.}\sim\bar\tau^{\dot\alpha}$ present, as it is the case when for instance the restriction of  $L_Q$  to $D$ is trivial, then they cannot be lifted by  (supersymmetric) bulk or world-volume fluxes.


\section{Non-perturbative superpotential and supersymmetry breaking fluxes}
\label{sec:SB}

If the bulk is characterized by a non-vanishing $G_{0,3}$-flux  the bulk is not supersymmetric and then we would expect several of the above arguments based on supersymmetry not to be valid anymore. On the other hand, taking a large universal modulus approximation  has the effect of diluting the fluxes and then the supersymmetry breaking  can be made small. Actually, in this regime the flux-induced supersymmetry breaking can be recovered  as a no-scale supersymmetry breaking vacuum in the low-energy effective action. Hence, it is instructive to revisit our results from the viewpoint of the low-energy interpretation.

In the small $\varepsilon$ approximation, it is reasonable to consider the warping to be approximately constant. Hence, we can reabsorb it in the internal K\"ahler metric by substituting $\d s^2_X\rightarrow e^{2A}\d s^2_X$, so that the universal modulus becomes the usual breathing  K\"ahler mode parametrizing the overall volume of the internal space. Let us take a basis of divisors $D^a$, $a=1,\ldots, h^{1,1}(X)$. Then an explicit basis of complexified K\"ahler moduli is given by 
\be
t^a=-\frac12\int_{D^a} J\wedge J-\ii\int_{D^a}  \tilde C_{\it 4}
\ee
where $\tilde C_{\it 4}$ is the modified SL$(2,\mathbb{Z})$ invariant R-R potential introduced in (\ref{invRR4}). Notice that these do not span the complete set of chiral fields, above denoted by $t^I$, which can appear in the expansion of the E3-bosonic action. More precisely, there are additional  moduli associated with the zero modes of the potentials $B_{\it 2}$ and $C_{\it 2}$. Let us assume for simplicity that they are absent.\footnote{In the orientifold limit, these are the moduli counted by $h^{1,1}_-(\tilde X)$, where $\tilde X$ is the double-cover CY three-fold. Hence, we assume that $h^{1,1}_-(\tilde X)=0$.}

The associated K\"ahler potential is given by (see for instance \cite{Denef:2008wq}):
\be
K=-2\log V_X
\ee
where $V_X$ denotes the volume of the internal space
\be
V_X=\frac{1}{3!}\int_X J\wedge J\wedge J
\ee
 which is now free to dynamically change.
In order to make the dependence of $K$ on the K\"ahler moduli $t^a$ more explicitly, let us work in cohomology and parametrize the K\"ahler form as $J=\xi_a[D^a]$ where $[D^a]$ denotes the closed two-form which is Poincar\'e dual to $D^a$ and $\xi_a$ provide K\"ahler moduli dual to the $\Re t^a$'s.\footnote{In our conventions Poincar\'e duality reads $\int_X\omega_{\it 4}\wedge [D^a]=-\int_{D^a}\omega_{\it 4}$ for any closed form $\omega_{\it 4}$.} We can then write
\be
V_X=\frac{1}{3!}\, d^{abc}\,\xi_a\,\xi_b\,\xi_c
\ee
where $d^{abc}$ is the triple intersection number $d^{abc}\equiv D^a\cdot D^b\cdot D^c$. Furthermore 
\be\label{txi}
\Re t^a=\frac12\, d^{abc}\,\xi_b\,\xi_c=\frac{\del V_X}{\del \xi_a}
\ee
The relation (\ref{txi}) provides the implicit dependence of $V_X$ and hence of $K$ on the $t^a$'s.  
On the other hand, the low-energy  superpotential takes the form \cite{Gukov:1999ya,GKP}
\be
W_{\rm tree}=\int_X\Omega\wedge G_{\it 3}
\ee
which does not depend on the K\"ahler moduli. One can show that $K^{a\bar b}K_a K_{\bar b}=3$, where $K_a\equiv\del K/\del t^a$ etc., and $K^{a\bar b}$ is the inverse of the K\"ahler metric $K_{a\bar b}$. Then the K\"ahler potential is of no-scale type,  the four-dimensional classical potential is positive definite and there is supersymmetry breaking if $\langle W_{\rm tree}\rangle \equiv W_0\neq 0$, that is if $G_{0,3}\neq 0$.

By decomposing the divisor $D$ wrapped by the E3-brane in the basis $D^a$, $D=n_a D^a$, we can write  the associated chiral field as
\be
t\equiv S_{\rm E3}=2\pi\, n_a t^a+...
\ee
where $...$ corresponds to terms containing $\calf$, $B_{\it 2}$ and $C_{\it 2}$, which do not scale with the volume.
Hence the associated F-term is given by $\langle F_T\rangle=2\pi n_a \langle F_T^a\rangle$ with
\be
 \langle F_T^a\rangle= \,c\,e^{\frac{K}{2}}\, K^{a\bar b}D_{\bar b}\bar W_{\rm tree}=\, c\,e^{\frac{K}{2}}\, K^{a\bar b}K_{\bar b}\,\bar W_0
\ee
where $c$ is some constant which parametrizes additional contributions not  coming form the K\"ahler moduli sector. 
One can evaluate $K^{a\bar b}K_{\bar b}$ by noticing that $V_X$ is homogeneous of degree 3 in $\xi_a$, which are in turn homogeneous of degree $1/2$ in $\Re t^a$. Hence $K_{a}=-\frac{1}{V_X}\frac{\del V_X}{\del \Re t^a}=-\xi_a/V_X$ is homogeneous of degree $-1$ in $\Re t^a$ so that
\be
K_{\bar a b}\,\Re  t^{ b}=\frac12 \frac{\del K_{\bar a}}{\del\Re t^{b}}\,\Re t^{b}= -\frac12 K_{\bar a}
\ee
and then $K^{a\bar b}K_{\bar b}=-2\Re t^a$. We then see that, up to a numerical factor,
\be\label{Fterm}
\langle F_T\rangle\sim \frac{n_a \Re t^a}{V_X}\, \bar W_0
\ee
By using (\ref{expS}), we can apply this result to write down the explicit form of the flux-induced term $\langle F_T\rangle \theta^2$, appearing in the  effective action for the E3-brane universal zero modes, as predicted by the four-dimensional effective viewpoint.

Let us now check that this term fits in with the microscopical viewpoint provided by the E3-brane effective action. Take the  first term in (\ref{SBaction}) and remember that we have to rescale the bulk K\"ahler metric by $\d s^2_X\rightarrow e^{2A}\d s^2_X$, so that  $\sqrt{\det h}\rightarrow e^{4A}\sqrt{\det h}$ and $\bar{G_{0,3}}\cdot\bar\Omega\rightarrow e^{-6A}\bar{G_{0,3}}\cdot\bar\Omega$.
We now make the reasonable assumption that $\bar{G_{0,3}}\cdot\bar\Omega$ is almost constant, so that it can then be replaced by its average: 
\be
\bar{G_{0,3}}\cdot\bar\Omega \,\simeq\, \frac{\int_X *_X\bar{G_{0,3}}\wedge\bar\Omega}{V_X}\, =\frac{\ii\int_X \bar\Omega\wedge \bar G_{\it 3}}{V_X}\, =\,\frac{\ii\,\bar W_0}{V_X}
\ee
Then If we substitute $\lambda\rightarrow \theta$ (with constant $\theta$) in (\ref{SBaction}) and recall that $\int_D\d^4\sigma\sqrt{\det h}=-\frac12\int_D J\wedge J=n_a\Re t^a$,  we get a term  which is exactly of the form $\langle F_T\rangle \theta^2$ with $ \langle F_T\rangle$ as in (\ref{Fterm}).  This supports the agreement between the microscopic and effective macroscopic  viewpoints.

Regarding the other fermionic zero modes,  the second term in (\ref{SBaction})  is of order $\calo(\varepsilon^3)$ and then is negligible. In particular, in the large volume regime, it cannot be used to generate a  lift of $\tilde\lambda^{\rm z.m.}\sim \bar\tau^{\dot\alpha}$ zero modes  which would provide a superpotential in a  supersymmetric low-energy effective theory. 
Furthermore, notice that, remarkably, the non-supersymmetric fluxes do not alter at all the remaining spectrum of fermionic zero modes
and so, in summary, the only leading order effect  on the zero-mode counting is on the goldstini $\theta^\alpha$, as expected from low-energy supersymmetry arguments. 

All these observations lead to the following conclusion:
under reasonable assumptions, in the large universal modulus regime we can consider just the effect of the supersymmetric component of the fluxes and write down the possible corresponding non-perturbative superpotential, while the effects of the non-supersymmetric component is already taken into account by the low-energy effective dynamics.\footnote{An alternative ten-dimensional approach to consistently incorporate the supersymmetry breaking $G_{0,3}$ flux in the instantonic calculations was suggested in \cite{Lust:2005cu}, in which it was proposed that a proper counting of the fermionic zero-modes on the E3-brane  should take into account the interaction of the E3-brane with the backreaction that it causes on the background fluxes and the geometry.}

Recall that in order to simplify the above discussion we omitted the subleading terms of the E3-brane bosonic action (\ref{E3action3}) which contain the world-volume fluxes and the bulk potentials $B_{\it 2}$ and $C_{\it 2}$, assuming the absence of the associated moduli. A more complete analysis should include these ingredients, but we expect the final conclusion to remain unchanged.   On the other hand, in particular regimes other subleading terms which we have considered negligible, as for instance the second term in (\ref{SBaction}), could become relevant and should be included as well.  It would be interesting to better explore the cases in which our assumptions are violated  and see whether the action (\ref{SBaction}) can be used to derive  physical effects which are not captured by the low-energy effective theory.


\section{Uplift to M5-brane on CY four-folds}
\label{sec:uplift}

So far we have discussed F-theory compactifications adopting the viewpoint provided by type IIB supergravity. This approach has allowed us to use directly the standard Green-Schwarz D3-brane effective action and furthermore is well suited to be compared and completed by perturbative string techniques.

On the other hand, several physical ingredients, and in particular the topological ones, can be 
given a more satisfying geometrical interpretation using the dual description provided by M-theory. The dual bulk configurations are given by M-theory warped flux compactifications \cite{Becker:1996gj,Dasgupta:1999ss} to $\mathbb{R}^{1,2}$ on a CY four-folds $Y$. $Y$ which is a holomorphic fibration of an elliptic curve $T^2_\tau$ over the three-fold $X$, where the IIB axio-dilaton $\tau$ becomes the complex structure of the elliptic fiber.

In the M-theory picture the E3-brane becomes a `vertical' Euclidean M5-brane, which wraps the divisor $D\subset X$ as well as the fiber $T^2_\tau$. The aim of this section is to uplift our fermionic effective action for the E3-brane to the one for vertical M5-branes. Hence, by covariance, we will actually obtain the fermionic effective action for Euclidean M5-branes on more general CY four-folds. 

Let us start by recalling how the M-theory and IIB bulk configurations are related, following the discussion of \cite{Denef:2008wq}. The M-theory metric takes the form
\be\label{11dmetric}
\d s^2_{11}=e^{\frac{8A}{3}}\d s^2_{\mathbb{R}^{1,2}}+e^{-\frac{4A}{3}}\d s^2_Y
\ee 
where $\d s^2_{\mathbb{R}^{1,2}}$ is the usual flat metric on ${\mathbb{R}^{1,2}}$ and $\d s^2_Y$ is the Ricci-flat CY metric on the elliptically fibered $Y$. Locally, if $\tau$ is slowly varying, $\d s^2_Y$ can be related to the type IIB metric by the following {\em approximate} split
\be\label{factform}
\d s^2_{Y}\simeq \d s^2_X+\d s^2_{T_\tau^2}\qquad~~~\text{with}\quad~~ \d s^2_{T_\tau^2}=\frac{L^2}{\Im\tau}\big[(\d x+\Re\tau\, \d y)^2+(\Im\tau)^2\d y^2\big]
\ee 
Here $L^2$ denotes the volume of the torus fiber as measured by the CY metric, while $x$ and $y$ are dimensionless coordinates with periodicities $x\simeq x+1$ and $y\simeq y+1$.

We recall that the duality between  these M-theory backgrounds and the IIB F-theory configurations is obtained by first reducing to type IIA along one circle of $T_{\tau}$ and then T-dualizing to type IIB along the second circle of $T_{\tau}$. Hence, the final duality  actually requires the strict limit $L\rightarrow 0$, in which the second circle decompactifies in type IIB and enhance the flat external directions from $\mathbb{R}^{1,2}$ to $\mathbb{R}^{1,3}$. 
 Although (\ref{factform}) is not the actual CY eight-dimensional metric (for $\tau$ holomorphically non-constant), 
 it is  nevertheless expected to provide a good description in the dual IIB background, since corrections should correspond to massive states which completely decouple in the $L\rightarrow 0$ limit. In particular, in this limit $e^A$ appearing in (\ref{11dmetric}) depends just on base $X$ and coincides with the IIB warping. In the following we will mostly work with local expressions and, in practice, we will  approximate $\tau$ to a constant. General arguments of covariance will then be invoked to extend the results to non-constant $\tau$ case. 
 
 The M-theory four-form flux $G_{\it 4}$ is related to the IIB three-form fluxes by
\be\label{G4M}
G_{\it 4}=L(-H_{\it 3}\wedge \d x +\tilde F_{\it 3}\wedge \d y)= e^{\frac{\phi}{2}}\Im(\bar G_{\it 3}\wedge \chi)
\ee
where we have introduced the $(1,0)$ vielbein along $T^2_\tau$:
\be
\chi=\frac{L}{\sqrt{\Im\tau}}(\d x+\tau\d y)
\ee
Supersymmetry requires that $G_{\it 4}$ is (2,2) and primitive, in agreement with the IIB prescription. On the other hand, the braking of supersymmetry is associated to a non-vanishing $G_{0,4}=\bar{G_{4,0}}$ component.

The modular group of the torus $T^2_\tau$ is given by the large coordinate transformations which preserve the coordinates identifications, namely: 
\be
\left(\begin{array}{c} 
x \\ y
\end{array}\right)  \quad
\rightarrow\quad\left(\begin{array}{cc} 
a & -b \\
-c & d
\end{array}\right) \left(\begin{array}{c} 
x \\ y
\end{array}\right)  \qquad~~\text{with}\quad~~~\left(\begin{array}{cc} 
a & -b \\
-c & d
\end{array}\right) \in \text{SL}(2,\mathbb{Z})
\ee
These act on $\tau$ as in (\ref{tauduality}) and provide the geometrization of the type IIB SL(2,$\mathbb{Z}$).
It is easy to see that $\chi$ transforms as an object with U(1)$_Q$ charge $q_Q=-1$ under a modular transformation and by requiring the invariance of $G_{\it 4}$ one obtains that the three-form fluxes must transform in agreement with the IIB prescription.  Furthermore, the CY 
 holomorphic (4,0)-form $\Omega_Y$ is related to the holomorphic (3,0)-form $\Omega_X$ on $X$ (we have added a suffix $_X$ for clarity) as follows
\be
\Omega_{Y}=L\,\Omega_X\wedge (\d x+\tau\d y)=e^{-\frac{\phi}{2}}\Omega_X\wedge\chi
\ee
Invariance of $\Omega_{Y}$ under $\text{SL}(2,\mathbb{Z})$ requires that $e^{-\frac{\phi}{2}}\Omega_X$ has $q_Q=1$, as it should.

Let us now pass to the vertical M5-brane. It wraps the vertical divisor $\hat D$ given by the fibration of $T^2_\tau$ over  $D$. 
Again we locally write $\hat D\simeq D\times T^2_\tau$, for $\tau$ approximately constant, and use the metric (\ref{factform}). 
In principle, one could go through the explicit chain of dualities relating the M5-brane to the E3-brane. However, in order to reconstruct the M5-brane world-volume quantities in terms of the E3-brane ones, we will follow the shortcut of  just requiring consistency with the bulk duality and with the relation between SL(2,$\mathbb{Z}$)-duality on the E3-brane and diffeomorphism invariance on the M5-brane. Hence, one can easily recognize that the E3-brane flux 
can be uplifted to a three-form flux $T_{\it 3}$ on the M5-brane as follows
\be
T_{\it 3}=L(-\calf\wedge \d x +\calf_{\rm D}\wedge \d y)=- e^{-\frac{\phi}{2}}\calf\wedge \chi
\ee
where in the last equality we have used the identity $\calf_{\rm D}=-\tau\calf$, which is valid for supersymmetric configurations.
Notice that the supersymmetric $T_{\it 3}$ is  (2,1) and primitive and then it is imaginary self-dual, $*_{\hat D}T_{\it 3}=\ii T_{\it 3}$.  
Furthermore, from the E3 Bianchi identities (\ref{covBI}), we obtain the M5 Bianchi identities
\be\label{M5bianchi}
\delbar T_{\it 3}=-\frac{\ii}{2}\, e^{\frac{\phi}{2}}\,\bar G_{\it 3}|_D\wedge \chi\quad~~~~~  \del T_{\it 3}=0
\ee
By recalling (\ref{G4M}) we see that this is not of the expected form $\d T_3= G_4|_{\hat D}$. In other words, part of the Bianchi identities/equations of motion is not generically fulfilled.  We have already observed this effect while discussing the E3-branes Bianchi's and equations of motion -- see footnote \ref{foot:BI} -- and (\ref{M5bianchi}) provide the M5-brane counterpart of it. It would be interesting to understand this point directly from the M-theory viewpoint.

\subsection{Uplift of the fermionic action}

Let us now pass to the fermionic sector, first turning off all bulk and world-volume fluxes.
In this case, already from the analysis of \cite{Witten:1996bn}, we know that the fermionic spectrum on the M5-brane can be expressed in terms of $(p,0)$ and $(0,q)$ forms $\Phi_{\it 0}\equiv\Phi_{(0,0)},\Phi_{\it 1}\equiv\Phi_{(0,1)},\Phi_{\it 2}\equiv\Phi_{(2,0)},\Phi_{\it 3}\equiv\Phi_{(0,3)}$ which take value in the rank-two spin-bundle along the external (Wick rotated) $\mathbb{R}^3$. One easily can realize that these can be related to the world-volume fermions on the E3-brane by a decomposition of the form:
\be\label{M5dec}
\begin{aligned}
\Phi_{\it 0}&=\lambda\\
\Phi_{\it 1}&=\psi-\frac{\ii}{2}\,\tilde\lambda\wedge\bar\chi\\
\Phi_{\it 2}&=-\rho+\ii\,\tilde\psi\wedge\chi\\
\Phi_{\it 3}&=\frac{\ii }{2}\,\tilde\rho\wedge\bar\chi\\
\end{aligned}
\ee 
The different coefficients are fixed by consistency of the following formulas. Notice that indeed $\tilde\lambda,\tilde\psi,\tilde\rho$ must have U(1)$_Q$ charges $q_Q=-1,+1,-1$ respectively.\footnote{In this decomposition the rank-two spinorial index associated with the normal $\mathbb{R}^3$  directions to the M5-brane 
 have to be interpreted as either left- or right-handed Weyl spinorial indices along  the  normal $\mathbb{R}^4$  directions to the E3-brane.
 This effect is due to the non-uniform appearance of the $\chi$ or $\bar\chi$ in the decomposition, which contains the additional flat direction of the type IIB picture and, after performing the chain of dualities relating the two descriptions,  is eventually responsible for the appearance of the  different four-dimensional chiralities.}

By using the decomposition (\ref{M5dec}), still assuming constant $\tau$,  it is easy to see that the E3 fermionic action  in the absence of fluxes is reproduced by the dimensional reduction of the following action for the M5-brane fermions:
\be\label{fluxlessM5}
2\int_{\hat D}(*\Phi_{\it 1}\wedge \del\Phi_{\it 0}+\,*\Phi_{\it 2}\wedge \delbar\Phi_{\it 1}+\,*\Phi_{\it 3}\wedge\del\Phi_{\it 2})
\ee
where $*$ is computed by using the bulk induced K\"ahler metric and, for instance, by $*\Phi_{\it 1}\wedge \del\Phi_{\it 0}$ we mean $\ii\epsilon_{\alpha\beta}*\Phi^\alpha_{\it 1}\wedge \del\Phi^\beta_{\it 0}$, where $\alpha,\beta$ are the three-dimensional sponsorial indices along $\mathbb{R}^3$.
Even though the action (\ref{fluxlessM5}) is derived for an M5-brane wrapping a vertical divisor $\hat D$ in a CY with factorized metric  (\ref{factform}) and constant $\tau$, by covariance it can be clearly extended to more general divisors $\hat D$ on more general CY four-folds. As a simple check, the equations of motion for the fermions simply require the zero modes to be given by harmonic forms (for instance $\delbar\Phi_{\it 1}=\delbar^\dagger\Phi_{\it 1}=0$), in agreement with \cite{Witten:1996bn}.


We can now proceed and include fluxes as well. As we have seen in the above sections, the E3-brane effective action contains some non-manifestly SL(2,$Z$) invariant terms, the $\tilde\psi\tilde\psi$ ones. These acquire a clearer structure once expanded in the large universal modulus limit which, in practice, can be traded for an expansion  in different powers of the world-volume and background fluxes.   Hence, in the following we consider uplift of the E3-brane fermionic action order by order in the fluxes:
\be
S^{\rm ferm}_{\text{M5}}=S^{\rm ferm}_{\text{M5}(0)}+S^{\rm ferm}_{\text{M5}(1)}+S^{\rm ferm}_{\text{M5} (2)}+S^{\rm ferm}_{\text{M5}(3)}+\ldots
\ee
 Let us stress again that, although we will determine these terms  under slightly restricted conditions -- namely factorized metric (\ref{factform}), constant $\tau$ and vertical divisor $\hat D$ -- by covariance they must be valid for the most generic divisors $\hat D$ and bulk CY four-folds.

First of all, the zero-th order terms are just  the kinetic terms modified by the warping:
\be
\label{warpedM5}
S^{\rm ferm}_{\text{M5}\rm (0)}=2\int_{\hat D}(*\Phi_{\it 1}\wedge \del^+\Phi_{\it 0}+\,*\Phi_{\it 2}\wedge \delbar^-\Phi_{\it 1}+\,*\Phi_{\it 3}\wedge\del^+\Phi_{\it 2})
\ee
where, as in the E3-brane action, $\del^+=\del+\del A$ and $\delbar^-=\delbar-\delbar A$.

Going to first order in the fluxes, we get
\be\label{fluxM5ferm}
S^{\rm ferm}_{\text{M5} (1)}=\frac12 \int_{\hat D}e^{2A}\left(* \Phi_{\it 3}\wedge \mathbf{S}\cdot\Phi_{\it 3}+ *\Phi_{\it 2}\wedge{\mathbf{R}}\cdot\Phi_{\it 2}\right)
\ee
where we have introduced the operators 
\be
\mathbf{S}: \Lambda^{0,3}_{\hat D}\rightarrow \Lambda^{3,0}_{\hat D}\quad~~~~~{\bf R}: \Lambda^{2,0}_{\hat D}\rightarrow \Lambda^{0,2}_{\hat D}
\ee
which act as follows:
\be
 \mathbf{S}\cdot\Phi_{\it 3}= \frac{1}{3!3!}\mathbf{S}_{abc}{}^{\bar {de f}}\Phi_{\bar{def}}\,\d s^a\wedge\d s^b\wedge \d s^c \quad~~~~
 {\mathbf{R}}\cdot\Phi_{\it 2}= \frac{1}{2!2!}{\mathbf{R}}_{\bar {a b}}{}^{cd} \Phi_{{cd}}\,\d\bar s^{\bar a}\wedge\d\bar s^{\bar b}
 \ee
with
\begin{subequations}
\begin{align}
 \mathbf{S}_{abc}{}^{\bar {de f}} &\equiv \frac{3}{2} (\Omega_Y)_{{\bf v}abc} \calk^{\bf v}{}_{g}{}^{[\bar d} T^{\bar {e f}]g} \\
 {\mathbf{R}}_{\bar {a b}}{}^{cd} &\equiv \frac12(\bar\Omega_Y)_{\bar {a b}}{}^{{\bf v}f}\Big[ \calk_{{\bf v}}{}^{g[c} T^{d]}{}_{gf} - \frac12G^{cd}{}_{{{\bf v}f}}\Big]\label{bfR}
\end{align}
\end{subequations}
Notice that if we relax the condition that $\tau$ is constant and compare  the r.h.s.\ of (\ref{bfR}) with the right-hand side of (\ref{Rexp1}), we can argue for the following identification 
\be\label{phident}
\calk_{{\bf v}}{}^{{\chi\chi}} = 2 {\bf v}(\phi)
\ee

At second order in the fluxes we get:
\be\label{fluxM5ferm2}
S^{\rm ferm}_{\text{M5} (2)}= \int_{\hat D}e^{4A}\,*\Phi_{\it 2}\wedge {\bf U}\cdot \Phi_{\it 3}
\ee
where we have introduced the operator 
\be
\mathbf{U}: \Lambda^{0,3}_{\hat D}\longrightarrow \Lambda^{0,2}_{\hat D}
\ee
defined by
\be
\mathbf{U}^{\bar a}= -\frac12\, T_{bc\bar d}\,(G_{\it 4})^{bc\bar{d a}}
\ee

The last terms we explicitly consider are the third order ones. By using (\ref{phident}), propose  the following form of such terms:
\be\label{fluxM5ferm3}
S^{\rm ferm}_{\text{M5} (3)}= \frac12 \int_{\hat D}e^{6A}\left(*\Phi_{\it 3}\wedge{\mathbf{S}'}\cdot\Phi_{\it 3}+*\Phi_{\it 2}\wedge{\mathbf{R}'}\cdot\Phi_{\it 2}\right)
\ee
where the operator $\mathbf{S}': \Lambda^{0,3}_{\hat D}\longrightarrow \Lambda^{3,0}_{\hat D}$ is given by
\begin{subequations}
\begin{align}
\mathbf{S}'_{abc}{}^{\bar{def}}&\equiv\frac3{4} (\iota_{a}T_{\it 3}\cdot \iota_g T_{\it 3})(G_{\it 4})^{{\bf v}g}{}_{bc}(\Omega_Y)_{{\bf v}}{}^{\bar{def}}\\
\mathbf{R}'_{\bar{ab}}{}^{{cd}}&\equiv\frac12 \delta^{\bar\chi}_{[\bar a}(\bar\Omega_Y)_{\bar b]}{}^{ef{\bf v}}\delta^{[c}_{\chi}(T_{\it 3})_{ef}{}^{d]} \big[{\bf v}(e^{-4A})-\frac14\calk_{{\bf v}}{}^{gh}(\iota_gT_{\it 3}\cdot \iota_h T_{\it 3})+T_{\it 3}\cdot\iota_{\bf v} G_{\it 4}\big]
\end{align}
\end{subequations}
We see that $\mathbf{R}'$ contains the explicit dependence on the directions $\chi$ and $\bar\chi$, and hence breaks the diffeomorphism invariance of the M5-brane action. This is however expected from the fact that no known diffeomorphism invariant action of the M5-brane, with minimal field content, is known. In a diffeomorphism invariant formulation \`a la \cite{Pasti:1997gx,Bandos:1997ui}, $\chi$ should be promoted to an auxiliary non gauge-fixed one-form.  The operator ${\bf R}'$ corresponds to the uplift of the operator $\tilde\calr'$ (\ref{calrprime}) on the E3-brane and, indeed, the fact that  ${\bf R}'$ cannot be put in a diffeomorphism invariant form is the M-theory counterpart  of the   non-SL(2,$\mathbb{Z}$)-invariance of  $\tilde\calr'$. Let us stress again that this operator is non-vanishing because of the off-shellness of the supersymmetric M5-brane configuration, in agreement with the general problem of writing an off-shell diffeomorphism invariant action for the M5-brane.

Finally, we can also uplift the E3-terms (\ref{SBaction}) containing a possible type IIB supersymmetry breaking component $G_{0,3}\neq 0$, which corresponds to  a non-vanishing  $G_{4,0}=\bar{G_{0,4}}$ in M-theory. The fermionic terms (\ref{SBaction}) uplift to
\be\label{M5SB}
\Delta^{\rm SB}S^{\rm ferm}_{\rm M5}=-\frac18\int_{\hat D}e^{2A}\, \bar\Omega_Y\cdot G_{\it 4}\, \Phi_{\it 0}\Phi_{\it 0}*1
-\frac{\ii}{16}\int_{\hat D}e^{6A}\,\Omega_Y\cdot G_{\it 4}\, (\Phi_{\it 1})^a\iota_a T_{\it 3}\wedge \Phi_{\it 1}\wedge T_{\it 3}
\ee

\subsection{Implications for the zero-modes}

Let us now discuss the effect of the fluxes on zero modes. One can in principle determine it  order by order in the perturbative expansion available in large universal modulus regime, as we did in section \ref{sec:zm} while discussing the spectrum of right-handed fermionic zero modes on the E3-brane.
\footnote{In section \ref{sec:zm}, we have seen that the spectrum of  left-moving fermionic zero-modes on the E3-brane can be understood exactly, while in order to handle the zero-modes of the right-handed fermions  it is more natural to make a perturbative analysis in the large breathing mode regime. On the M5-brane there is no distinction between left-moving and right-moving fermions, since this is translated into the transformation property of the different components under the U(1) local rotation $\chi\rightarrow e^{\ii\alpha}\chi$ along the elliptic fiber $T^2_\tau$, which cannot be defined in the most general case.} 

As we have explained in section \ref{sec:pert}, the perturbative expansion is obtained by setting $e^{-4A}=\frac{1}{\varepsilon^2}+e^{-4\hat A}$, with $e^{-4\hat A}$ having a fixed normalization, and taking $\varepsilon$ very small. We will work up to order $\calo(\varepsilon)$.
Then, by using the fact that $e^{2A}\simeq \varepsilon[1+\calo(\varepsilon^2)]$ and $\d A\simeq \calo(\varepsilon^2)$, we see that we can restrict our attention on the terms 
(\ref{warpedM5}) and (\ref{fluxM5ferm}). In particular,  in (\ref{warpedM5}) we can replace $\del^+$ with $\del$ and $\delbar^-$ with $\delbar$ and in (\ref{fluxM5ferm}) we can replace $e^{2A}$ with $\varepsilon$.

It is then easy to see that at the zero-order in $\varepsilon$, the spectrum is the same as in the flux less case considered in \cite{Witten:1996bn}, namely: 
\be
\begin{aligned}
&\Phi^{\rm z.m.}_{{\it 0}(0)}\in \text{Harm}_{\del}^{0,0}(\hat D)\quad~~~~ \Phi^{\rm z.m.}_{{\it 1}(0)}\in\text{Harm}_{\delbar}^{0,1}(\hat D)\\
&\Phi^{\rm z.m.}_{{\it 2}(0)}\in \text{Harm}_{\del}^{2,0}(\hat D)\quad~~~~\Phi^{\rm z.m.}_{{\it 3}(0)}\in \text{Harm}_{\delbar}^{0,3}(\hat D)
\end{aligned}
\ee
We can now go to order $\varepsilon$ and by adapting  to the present case the discussion made for the E3-brane we see that the (supersymmetric) fluxes can lift the zero-modes  $\Phi^{\rm z.m.}_{{\it 2}(0)}$ and $\Phi^{\rm z.m.}_{{\it 3}(0)}$. More specifically, by defining the operators 
\begin{subequations}
\begin{align}
{\bf P}_{\it 2}&\equiv\calh_{\delbar}\circ{\bf R}\ :\ \text{Harm}_{\del}^{2,0}(\hat D)\rightarrow \text{Harm}_{\delbar}^{0,2}(\hat D)\\
{\bf P}_{\it 3}&\equiv\calh_{\del}\circ{\bf S}\ :\ \text{Harm}_{\delbar}^{0,3}(\hat D)\rightarrow \text{Harm}_{\del}^{3,0}(\hat D)
\end{align}
\end{subequations}
the surviving zero-modes are
\be\label{Phicond}
\Phi^{\rm z.m.}_{{\it 2}(0)}\in \ker {\bf P}_{\it 2}\quad~~~~~~~~ \Phi^{\rm z.m.}_{{\it 3}(0)}\in \ker {\bf P}_{\it 3}
\ee

On the other hand, the zero modes $\Phi^{\rm z.m.}_{{\it 0}(0)}$ and $\Phi^{\rm z.m.}_{{\it 1}(0)}$ remain untouched and it is interesting to observe that, by looking at (\ref{fluxM5ferm2}) and (\ref{fluxM5ferm3}),  this conclusion remains valid at all orders in $\varepsilon$. 

Notice that by forcing $T_{\it 3}\equiv 0$ the operator ${\bf P}_{\it 3}$ becomes trivial and,  at order $\calo(\varepsilon)$, one is left with the residual non-trivial condition $\Phi^{\rm z.m.}_{{\it 2}(0)}\in \ker {\bf P}_{\it 2}|_{T_{\it 3}=0}$, which agrees with what found in \cite{Saulina:2005ve,Kallosh:2005gs}. 

 We can also give a clear geometrical interpretation of the possible lift of the zero modes $\Phi^{\rm z.m.}_{{\it 3}(0)}$ provided by the condition (\ref{Phicond}). The logic is the same followed for the E3-brane in section \ref{sec:geomint} and so we will be sketchy.  The normal vectors  $\Omega^{\bar{{\bf v}abc}}\Phi^{\rm z.m.}_{(0)\bar{abc}}$ describe the possible infinitesimal geometrical deformation of the M5-brane which preserve the holomorphy of the embedding $\hat D$.
 On the other hand, such a deformation can change the complex structure on $\hat D$ and this in turn can change the nature of the world-volume flux, which can acquire a (3,0) component which cannot be compensated by any deformation of the world-volume two-form field. Such a component is exactly given by  ${\bf P}_{\it 3}\cdot \Phi^{\rm z.m.}_{{\it 3}(0)}$. Hence, if ${\bf P}_{\it 3}\cdot \Phi^{\rm z.m.}_{{\it 3}(0)}\neq 0$ then  the geometrical deformation described by $\Phi^{\rm z.m.}_{{\it 3}(0)}$ does violate the supersymmetry condition on the world-volume flux and then cannot be considered as a geometrical zero mode of the complete M5-brane configuration. 

Finally, regarding the terms (\ref{M5SB}) containing the supersymmetry breaking bulk flux, one may repeat  the discussion presented in section \ref{sec:SB} for the corresponding terms (\ref{SBaction}) on the dual E3-brane.

\section{Conclusions and future directions}

In this paper we have studied the effective action governing the (neutral) fermionic quantum fluctuations of Euclidean D3-brane, or E3-brane, instanton in F-theory flux compactifications. We have started from the (Wick-rotated)  supergravity Green-Schwarz formulation of the E3-brane effective action and worked at the quadratic order in the quantum fermionic fields but to all orders in the bulk fields and classical brane configuration. This is the starting point for one-loop semiclassical computations. One of our main results is the fermionic action presented in section \ref{sec:fermeff}  (and its uplift to the dual M5-brane action in section (\ref{sec:uplift}), in which the fermionic fields are topologically twisted along the world-volume directions and which encodes the complete dependence of the bulk  axio-dilaton, fluxes and warping, as well as on the world-volume flux. We have studied the spectrum of fermionic zero-modes, providing also a geometric interpretation of part of the zero mode lifting  mechanisms induced by the fluxes. We have explicitly considered the effect of a possibly non-vanishing supersymmetry breaking $(0,3)$ component of the $G_{\it 3}$-flux  and discussed in which sense it agrees with what expected from four-dimensional low-energy arguments. 

It could be useful to briefly summarize the qualitative effect of bulk and world-volume fluxes on the fermionic  zero mode structure, in the descriptions which are more commonly used in the literature: the M-theory description and the type IIB orientifold limit (see section \ref{sec:Oplanes}). In absence of bulk and world-volume fluxes the fermionic zero-modes are counted by  $h^{0,I}(\hat D)$ (with $I=0,1,2,3$)  and $h^{0,i}_{\pm}(\tilde D)$ (with $i=0,1,2$) on the M5-brane  and (double-cover) E3-brane respectively -- see \cite{Blumenhagen:2010ja,Bianchi:2011qh}  for detailed discussions on their relation. Then supersymmetric bulk and world-volume fluxes generically affect the zero-modes corresponding to 
$h^{0,2}(\hat D),h^{0,3}(\hat D)$ and $h^{0,1}_-(\tilde D),h^{0,2}_\pm(\tilde D)$ on the M5-brane and E3-brane respectively, while they do not affect the remaining zero modes counted by $h^{0,0}(\hat D),h^{0,1}(\hat D)$ and $h^{0,0}_\pm(\tilde D),h^{0,1}_+(\tilde D)$ respectively. In particular, we stress that the pseudo-universal zero modes $\bar\tau^{\dot\alpha}$  on the E3-brane, which are counted by $h^{0,0}_-(\tilde D)$, {\em cannot}  be lifted by fluxes, at least at the one-loop order we are working in. On the other hand, a bulk supersymmetry breaking flux couples to $h^{0,0}(\hat D),h^{0,1}(\hat D)$ and $h^{0,0}_\pm(\tilde D)$ respectively -- see section \ref{sec:SB} for a discussion on the interpretation and physical implications of these couplings.  

The implications of our findings on the effective four-dimensional theory have been discussed only schematically  and these aspects clearly deserve a deeper study. In particular it would be worth  looking for a systematic and more topological way to compute the effect of fluxes on the zero-mode spectrum. Furthermore, we have neither considered higher order fermionic interactions, which may lift zero-modes of the quadratic action, nor charged fermionic modes, which should be consistently incorporated in the picture to provide a more complete characterization of the possible terms appearing in the effective action. Another aspect which would require a more detailed study is the complete supersymmetric structure of the E3-brane effective theory, which for instance would be useful to obtain more quantitative information on the E3-brane partition function. This requires the inclusion of the bosonic quantum fluctuations in the discussion. This sector is typically problematic, because of the well known problems related to the self-duality of the two-form gauge field living on the M5-brane \cite{Witten:1996hc} (see also the recent \cite{Kerstan:2012cy}). Understanding the complete E3-brane/M5-brane effective theory describing both bosonic and fermionic quantum fluctuations should be instrumental to clarifying the issues with SL(2,$\mathbb{Z}$) symmetry discussed in section \ref{sec:offshell}. 
   
Clearly, the final goal would be to gain control over the complete global structure of the effective superpotential, which is necessary  to  unambiguously determine  the vacuum structure of flux compactifications and their low-energy  effective action.

\vspace{2cm}

\centerline{\large\em Acknowledgments}

\vspace{0.5cm}

\noindent We would like to thank A.~Collinucci, F.~Fucito, F.~Marchesano, J.F.~Morales, D.~Sorokin and A.~Uranga for useful discussions.
M.~B.~would like to express his gratitude to A.~Hanany and the other members of the Theory Group at Imperial College for their kind hospitality and support while this work was being completed. This work was partially supported by the Italian MIUR-PRIN contract 2009KHZKRX-007 and by the ERC Advanced Grant n. 226455.

\vspace{3cm}

\newpage

\vspace{0.5cm}

\begin{appendix}

\section{Notations and conventions}
\label{app:conv}

We use the following sets of indices:
\begin{center}
\begin{tabular}{cl}
$M,N,\ldots$ & ten-dimensional spacetime\\
$m,n,\ldots$ & internal six-dimensional space $X$\\
$i,j,\ldots$ ($\bar\imath,\bar\jmath,\ldots$) & (anti)holomorphic indices on $X$\\
$A,B,\ldots$ & four-dimensional indices along $D$\\
$a,b,\ldots$ ($\bar a,\bar b$,\ldots) & (anti)holomorphic indices along $D$
\end{tabular}
\end{center}
Underlined indices are flattened using the appropriate vielbein. Tensor components along the normal bundles $N_D^{\rm 1,0}$, $N_D^{\rm 0,1}$ are denoted with indices $\bf v$, $\bar{\bf v}$ respectively.

The Hodge star in a $d$ dimensional space reads:
\begin{equation}
 \ast_{d} \omega_p =
 \frac{1}{p!(d-p)!} \epsilon_{\underline M_1\ldots \underline M_{d}}\,
 \omega^{\underline M_{d-p+1}\ldots \underline M_{d}}\, e_{d}^{\underline M_1}\wedge\ldots\wedge e_{d}^{\underline M_p}
\end{equation}
where $\epsilon_{0\ldots 9}=+1$ for the ten dimensional space-time, $\epsilon_{1\ldots6}=+1$ and $\epsilon_{1\ldots4}=+1$ for the Levi--Civita symbols on $X$ and $D$. Notice that $*_X$ is the Hodge star on  the internal space $X$ defined using the six-dimensional K\"ahler metric. Similarly, the Hodge star on $D$ is defined using the pull-back of the K\"ahler metric.

We define the contraction of two forms as:
\begin{equation}
 \omega_{\it p} \cdot \chi_{\it q} \equiv \frac1{p!(q-p)!} \omega^{M_1\ldots M_p} \chi_{M_{1}\ldots M_p\, M_{p+1}\ldots M_q} \d z^{M_{p+1}}\wedge\ldots\wedge\d z^{M_q}
 \quad (p\le q)
\end{equation}
When we contract forms with only internal or world-volume indices, we use the K\"ahler metric on $X$ or its pull-back on $D$.

\section{Details of the computation}
\label{app:details}
In this appendix we show some details about the derivation of the action (\ref{fullaction}). 

Start from the general action (\ref{fermact}). By using the explicit form of the bulk structure described in section \ref{sec:bulksugra} and the gamma-matrices (\ref{def Gamma}),  
 the action of the operators \eqref{susy operators} on $\Theta$ can be written explicitly as:
\begin{subequations}\label{explicit susy operators}
\begin{align}
\cald_m \Theta &= \Big\{\nabla_m + e^{2A-\phi/2}\Big(\frac{1}{4} \iota_mH_{\it3} \sigma_3 + \frac{e^\phi}{8} F_{\it3} \Gamma_m \sigma_1\Big)
+\frac{1}{2}(1+\gamma_7\sigma_2)\d A\, \Gamma_m \cr
&\quad-\frac{1}{8}\left[\d\phi+(\delbar\phi-\del\phi)\sigma_2\right]\Gamma_m-\frac{1}{2}\del_m(A-\phi/4)\Big\} \Theta\\[.5em]
\calo\,\Theta &= e^{A-\frac\phi4}\Big[e^{2A-\frac{\phi}2}\Big(\frac{1}{2} H_{\it3} \sigma_3 -\frac{1}{2}e^{\phi} {F}_{\it3} \sigma_1\Big) +\d\phi +(\delbar\phi-\del\phi)\sigma_2\Big]\Theta
\end{align}
\end{subequations}
We used the fact that $\Gamma_{11}\Theta=\Theta$ and $F_{\it5} = -4\ii\, e^{-\phi} (1+\Gamma_{11})(\bbone_4 \otimes \gamma_7) \d A$.
The covariant derivative $\nabla_m$ is defined in terms of the K\"ahler metric on $X$, and all warping and dilaton factors are taken into account explicitly. Similarly, we define $\Gamma_m \equiv \Gamma_{\underline m} e^{\underline m}_m$ where $e^{\underline m}_m$ is the vielbein associated to the K\"ahler metric and $\Gamma_{\underline m}$ are defined in \eqref{def Gamma}.
World-volume indices are also raised and lowered with the pull-back of the K\"ahler metric on $D$.
We remind that contraction of forms with gamma-matrices is implicit, hence for instance in the expressions above $\iota_m H_{\it3} = \frac12 (H_{\it3})_{mnp} \Gamma^{np}$ and $\d A = \del_m A\, \Gamma^m$.

It is useful to rescale the bispinor as $\Theta \equiv e^{\frac{3A}{2} +\frac\phi8} \Theta'$, so that the Ansatz \eqref{Fsplit} reads:
\begin{subequations}\label{ansatz fermions}
\begin{align}
& \left\{ \begin{array}{l}
\Theta'{}_1^{\rm L}=\frac12\Big(\lambda\otimes \eta^*+(M^{-1})_B{}^A\psi_A\otimes \gamma^B\eta+\frac14\rho_{AB}\otimes \gamma^{AB}\eta^*\Big)
 \\
\Theta'{}_2^{\rm L}=\frac\ii2\Big(\lambda\otimes \eta^*-(M^{-1})^A{}_B\psi_A\otimes \gamma^B\eta+\frac{1}4\rho_{AB}\otimes \gamma^{AB}\eta^*\Big)
\end{array}  \right.\\
&\left\{ \begin{array}{l} 
\Theta'{}_1^{\rm R}=\frac\ii2\Big(\tilde\lambda\otimes \eta+(M^{-1})_B{}^A\tilde\psi_A\otimes \gamma^B\eta^*+\frac14\tilde\rho_{AB}\otimes \gamma^{AB}\eta \Big)
\\
\Theta'{}_2^{\rm R}=-\frac12\Big(\tilde\lambda\otimes \eta-(M^{-1})^A{}_B\tilde\psi_A\otimes \gamma^B\eta^*+\frac{1}4\tilde\rho_{AB}\otimes \gamma^{AB}\eta\Big)
\end{array}  \right. 
\end{align}
\end{subequations}
Then, since $\bar\Theta \Gamma_m \Theta = \bar\Theta\Gamma_m \sigma_3 \Theta = 0$, the overall warping and dilaton factors are brought in front of the expression \eqref{fermact} that we need to compute, which then takes the form:
\begin{align}\label{initial expression}
S^{\rm ferm} &= \ii\int\d^4\sigma\ \sqrt M \ \bar\Theta'\left[(M_\sigma^{-1})^{AB}\Gamma_A \cald_B - \frac12e^{-A+\phi/4}\calo\right]\Theta'
\end{align}
where $\bar\Theta' = \Theta'^T \Gamma^{\ul0}$ and $\sqrt M = \sqrt{\det(M_{AB})}$.

In order to guide the reader through the various steps of the computation, we will divide $S_{\rm tot}^{\rm ferm}$ in several pieces:
\begin{equation} 
S_{\rm tot}^{\rm ferm}= S_\nabla + S_{G_{\it 3}} + S_{A} + S_\tau
\end{equation}
where $S_\nabla$ contains all the contributions coming from the covariant derivative $\nabla_m$ in $\cald_m$, $S_{G_{\it 3}}$ is associated to all the terms where the $H_{\it 3}$ and $F_{\it 3}$ fluxes appear, and $S_A,\ S_\tau$ contain the further contributions coming from the terms in \eqref{explicit susy operators} with derivatives of the warping and of the axio-dilaton respectively.

\subsection{Kinetic terms}\label{sub:kinetic terms}
We begin with the computation of $S_\nabla$, containing the terms arising from the covariant derivative $\nabla_m$ in \eqref{explicit susy operators}. After several manipulations, we obtain the following expression:
\begin{align}\label{kinetic terms}
S_{\nabla}&= 2\int \d^4\sigma\ \sqrt h \left(\psi^a \del_a\lambda+\tilde\psi_a\bar\del^a_Q\tilde\lambda-\psi^b\bar\del^a\rho_{ab}-\tilde\psi_b\del_{Q\,b}\tilde\rho^{ab}\right)\\\notag
&\quad
-\frac18  \int \d^4\sigma\ \sqrt h\ e^{2A}\left(\tilde\rho^{ab}\tilde\rho^{cd}e^{-\phi} \calk^{\bf v}{}_{e[a}\calf_{b]}{}^e \,\Omega_{{\bf v}cd}
\rho_{ab}\rho_{cd} e^{-\phi}\calk_{\bf v}{}^{e[a}\calf^{b]}{}_e \,\bar\Omega^{{\bf v}cd}\right) + S_{\rm extra}
\end{align}
where  $\del_Q$, $\bar\del_Q$ are defined below \eqref{FactionR}, $h = \det(h_{AB})$ and $S_{\rm extra}$ contains contributions coming from integration by parts of the terms proportional to $\bar\del\psi, \del\tilde\psi$. In our general setting with world-volume flux and non-trivial warping and axio-dilaton we obtain many such contributions, and we need to keep track of them to eventually get to the correct result. 
Notice in particular that the derivatives in \eqref{kinetic terms} are still missing the modifications due to the warping, as defined below \eqref{FactionL}.
We find the extra terms:
\begin{equation}\label{int by part terms}
\begin{split}
S_{\rm extra}=&-2\int\d^4\sigma
 \sqrt{h}\left(
 \lambda\psi^a \partial_aA+\tilde\lambda\tilde\psi_a \partial^aA
-\psi^a\rho_{ab}\partial^bA-\tilde\psi_a\tilde\rho^{ab}\partial_bA \right)
\\
&+\int\d^4\sigma\ \tfrac{h}{\sqrt{M}}e^{4A-\phi}\left( 
-\tilde\lambda(\calf^2\tilde\psi)_a\partial^a(A-\phi/4) -\lambda(\calf^2\psi)^a\partial_a(A-\phi/4) \right)
\\
&+\int\d^4\sigma\ \tfrac{h}{\sqrt{M}}e^{4A-\phi}\left( 
(\calf^2\tilde\psi)_a\tilde\rho^{ab}\partial_b(A-\phi/4) +(\calf^2\psi)^a\rho_{ab}\partial^b(A-\phi/4) \right)
\\
&+\int\d^4\sigma\ \tfrac{h}{\sqrt{M}}e^{4A-\phi}\left(
-\frac12\tilde\lambda(\calf\tilde\psi)_a(\nabla\cdot\calf)^a -\frac12\lambda(\calf\psi)^a(\nabla\cdot\calf)_a \right)
\\
&+\int\d^4\sigma\ \tfrac{h}{\sqrt{M}}e^{4A-\phi}\left(
\frac12(\calf\tilde\psi)_a\tilde\rho^{ab}(\nabla\cdot\calf)_b +\frac12(\calf\psi)^a\rho_{ab}(\nabla\cdot\calf)^b
\right)
\end{split}
\end{equation}
where for brevity we set $(\calf\tilde\psi)_a = \calf_a{}^b \tilde\psi_b$, $(\calf^2\tilde\psi)_a = \calf_a{}^b \calf_b{}^c \tilde\psi_c$, and similarly for $\psi$.
Moreover, we write $(\nabla\cdot\calf)_A = \nabla^B \calf_{BA}$.

\subsection{Three-flux contributions}
We now begin to add the extra couplings coming from the operators $\cald_A$ and $\calo$. We start with $S_{G_{\it3}}$, containing the terms proportional to the bulk flux $G_3$.
We can write it explicitly as follows:
\begin{align}
S_{G_{\it3}}&=\frac\ii4\int \d^4\sigma\ e^{2A-\phi/2} \sqrt{ M}\Big(\bar\Theta_1 (M^{-1})^{AB} \Gamma_A \iota_B H_{\it3} \Theta_1 
- \bar\Theta_2 (M^{-1})^{BA} \Gamma_A \iota_B H_{\it3} \Theta_2\Big)\\\notag
 &\quad+\ii\int \d^4\sigma\ e^{2A-\phi/2} \sqrt{ M}\Big(
 \frac{e^\phi}{8}  (M^{-1})^{AB}\bar\Theta_1\Gamma_A  F_{\it3}  \Gamma_B \Theta_2
 +\frac{e^\phi}{4} \bar\Theta_1  F_{\it3}  \Theta_2-\frac{1}{4} \bar\Theta_1  H_{\it3}  \Theta_1\Big)\cr\notag
  &\quad+\ii\int \d^4\sigma\ e^{2A-\phi/2} \sqrt{ M}\Big(
 \frac{e^\phi}{8}(M^{-1})^{BA} \bar\Theta_2 \Gamma_A  F_{\it3}  \Gamma_B \Theta_1
 +\frac{e^\phi}{4} \bar\Theta_2  F_{\it3}  \Theta_1+\frac{1}{4} \bar\Theta_2  H_{\it3}  \Theta_2\Big)
\end{align}
The computation turns out to be easier if we use the imaginary self duality (ISD) property of $G_{\it 3}$ to split $H$ and $F_3$ in ISD ($+$) and IASD ($-$) parts:
\begin{equation}\label{isd and aisd splitting}
H_{\it3} ^+=-\ii e^{\phi}F_{\it 3}^+=H_{\it2,1}^{\rm P}+H_{\it0,3}\qquad
H_{\it3} ^-=\ii e^{\phi}F_{\it 3}^-=H_{\it1,2}^{\rm P}+H_{\it3,0}
\end{equation}
and at the same time we define combinations $\Theta_\pm\equiv\frac12(\Theta'_1\pm\ii\Theta'_2)$ of the kappa-fixed fermions with definite U(1)$_Q$ charge:
\begin{equation}\label{theta plus minus}
\begin{split}
\Theta^R_+&=\frac\ii2(M^{-1})^{(AB)}\tilde\psi_A\otimes\gamma_B\eta^*,\\[.4em]
\Theta^L_+&=\frac12(M^{-1})^{(AB)}\psi_A\otimes\gamma_B\eta,\\[.2em]
\Theta^R_-&=\frac\ii2\Big(\tilde\lambda\otimes\eta-(M^{-1})^{[AB]}\tilde\psi_A\otimes\gamma_B\eta^*+\frac14\tilde\rho_{AB}\otimes\gamma^{AB}\eta\Big),\\
\Theta^L_-&=\frac12\Big(\lambda\otimes\eta^*-(M^{-1})^{[AB]}\psi_A\otimes\gamma_B\eta+\frac14\rho_{AB}\otimes\gamma^{AB}\eta^*\Big).
\end{split}
\end{equation}
Then we can write: 
\begin{equation} \label{rearranged S G3}
\begin{split}
S_{\rm G_3}&=\ii\int\d^4\sigma\  \sqrt{h}\ e^{2A-\phi/2}\Big\{\big(1-\tfrac{\sqrt{M}}{\sqrt h}\big)\left(\bar\Theta_+  H_{\it3}^-\Theta_++\bar\Theta_-H_{\it3}^+\Theta_-\right)\\[.3em]
&\quad+ h^{AB}\bar\Theta_+\Gamma_A \,\iota_BH_{\it3}^+\Theta_+- \bar\Theta_+ H_{\it3}^+\Theta_++ h^{AB}\bar\Theta_-\Gamma_A \,\iota_BH_{\it3}^-\Theta_-- \bar\Theta_- H_{\it3}^-\Theta_- \\[.3em]
&\quad-e^{2A-\phi/2}\calf^{AB}\Big(\bar\Theta_+\Gamma_A \,\iota_BH_{\it3}^-\Theta_-- \frac14\bar\Theta_+\Gamma_{AB} H_{\it3}^-\Theta_-- \frac14\bar\Theta_+ H_{\it3}^-\Gamma_{AB}\Theta_- \Big)\\[.3em]
&\quad-e^{2A-\phi/2}\calf^{AB}\Big(\bar\Theta_-\Gamma_A \,\iota_BH_{\it3}^+\Theta_+- \frac14\bar\Theta_-\Gamma_{AB} H_{\it3}^+\Theta_+- \frac14\bar\Theta_- H_{\it3}^+\Gamma_{AB}\Theta_+ \Big)\ \Big\}
\end{split}
\end{equation}
The Left spinors give rise to the contributions:
\begin{equation}\label{S G3 left}
\begin{split}
S_{G_3}^{\rm Left} &=\frac14 \int\d^4\sigma\ \sqrt h\ e^{2A-\phi}
\Big(\lambda\lambda(H\cdot\bar\Omega)+\frac14\rho_{\bar{ab}}\rho_{\bar{cd}} \bar\Omega^{ab{\bf v}} H^{cd}{}_{\bf v}\Big)\cr
&\quad+4\int\d^4\sigma\ \tfrac{h}{\sqrt{M}}\ e^{4A-\phi} \Big(\psi^a \rho_{ab} (\calf\cdot H)^b
-\lambda\psi_{\bar a}(\calf\cdot H)^{\bar a}\Big),
\end{split}
\end{equation}
while the Right part can be cast in the  following form:
\begin{equation}\label{S G3 right}
\begin{split}
S_{G_3}^{\rm Right} &= \frac14\int\d^4\sigma\ e^{2A-\phi}
(\sqrt M-\sqrt h)\left[\tilde\lambda\tilde\lambda(H\cdot\Omega)+\frac14\tilde\rho_{\bar{ab}}\tilde\rho_{\bar{cd}} \Omega^{\bar{cd{\bf v}}} H^{\bar{ab}}{}_{\bar{\bf v}}\right]
\\
&\quad-\int\d^4\sigma\ e^{2A-\phi} \sqrt h\left[
\frac12\tilde\psi_a\tilde\psi_b \bar\Omega^{ac{\bf v}} H_{c{\bf v}}{}^b
+e^{2A} 2\tilde\rho_{\bar{ab}}\tilde\psi^{\bar a} \calf^{c\bar d} H_{c\bar d}{}^{\bar b}\right]\\
&\quad-\int\d^4\sigma\ e^{4A-\phi}\tfrac{h}{\sqrt M} \left[ \tilde\lambda\tilde\psi_a \calf^{c\bar d} H_{c\bar d}{}^a 
-\tilde\rho_{\bar{ab}}\tilde\psi^{\bar a} \calf^{c\bar d} H_{c\bar d}{}^{\bar b}\right]\\
&\quad+\frac14\int\d^4\sigma\ e^{6A-\phi} \tfrac{h}{\sqrt M} \calf^{a\bar b}(\ii G_{a\bar b {\bf v}})\bar\Omega^{ab{\bf v}}\tilde\psi_a(\calf\tilde\psi)_b
\end{split}
\end{equation}

\subsection{Extra contributions from a non-trivial warping}
Some terms promotional to $\d A$ were already present in $S_{\rm extra}$. However, derivatives of the warping appear explicitly in $\cald_m$ \eqref{explicit susy operators}, and we collect the resulting contributions in $S_A$.
The starting point is
\begin{equation}
\begin{split}
S_{A}&= \frac{\ii}2\int\d^4\sigma\ \sqrt M\ \bar\Theta' (M_\sigma^{-1})^{AB}\,
\Gamma_A\,\d A\, \Gamma_B\, (1+\gamma^7\sigma_2)\Theta'\\
&=\ii\int\d^4\sigma\ \sqrt h\ \ii\bar\Theta_2 (2\Gamma^A\partial_A A-4\d A)(\Theta_1^R -\Theta_1^L)\cr
&\quad+4\ii\int\d^4\sigma\ e^{2A-\phi/2}\sqrt h\ (\bar\Theta_-^R+\bar\Theta_+^L)\, \calf\, \d A (\Theta_-^R + \Theta_+^L)
\end{split}
\end{equation}
The result is:
\begin{align}\label{S F5}
S_{A}=\ &\int\d^4\sigma\,\tfrac{h}{\sqrt M} \left(2 \lambda\psi^a\partial_aA -2\psi^a \rho_{ab} \partial^bA\right)\\\notag
&-2\int\d^4\sigma\,\tfrac{h}{\sqrt M}\Big\{\tilde\lambda\left[(1-2e^{2A-\phi/2}\calf^2)\tilde\psi\right]_a\,\partial^aA - \left[(1-2e^{2A-\phi/2}\calf^2)\tilde\psi\right]_a\,\tilde\rho^{ab}\partial_bA\Big\}\notag
\end{align}
Notice that most of these terms combine with \eqref{int by part terms} in order to give the warping modifications to the kinetic terms as defined below \eqref{FactionL}.
In order to see this, the following identity turns out to be useful:
\begin{equation}\label{h and F identity}
h_{AB}-e^{4A-\phi}\calf_A{}^C\calf_{CB} = \sqrt{\frac{\det M}{\det h}} h_{AB}
\end{equation}

\subsection{Non-constant axio-dilaton}
Similarly to the warping contributions in the previous section, derivatives of the axio-dilaton appear explicitly in \eqref{explicit susy operators} and we collect their contributions in $S_\tau$. The resulting terms will eventually combine with \eqref{int by part terms} in the final result,
but we need to compute those arising from $F_1$ in $D_A$ and $\calo$.
After some algebra we can reduce $S_\tau$ to the following expression:
\begin{equation}
\begin{split}
S_{\tau} &= \frac{\ii}2 \int\d^4\sigma\ (\sqrt h -\sqrt M)\,\bar\Theta'\left((1-\sigma_2)\del\phi+(1+\sigma_2){\bar\del}\phi\right)\Theta'\\
&\quad-\frac{1}{4}\int\d^4\sigma\ \sqrt h\ \bar\Theta'\left[\Gamma^A\left((1-\sigma_2)\del_A\phi+(1+\sigma_2)\bar\del_A\phi\right)\right]\Theta'\\
&\quad-\frac{1}{4}\int\d^4\sigma\ \sqrt h\ e^{2A}\ \bar\Theta'\left[\sigma_3 \calf\left((1-\sigma_2)\del\phi+(1+\sigma_2){\bar\del}\phi\right) \right]\Theta'
\end{split}
\end{equation}
Using $(1\mp\sigma_2)\Theta' = 2(\Theta_{\pm},\mp\ii\Theta_{\pm})$, we can rewrite it in a form that greatly simplifies the computation:
\begin{align}
S_{\tau} &= \ii\int\d^4\sigma\ (\sqrt h -\sqrt M)\left(2\bar\Theta_-\del\phi\,\Theta_+ +2\bar\Theta_+{\bar\del}\phi\,\Theta_-\right)\\\notag
&\quad-\ii\int\d^4\sigma\ \sqrt h\Big[\left(\bar\Theta_-\Gamma^a\del_a\phi\,\Theta_+ +\bar\Theta_+\Gamma^{\bar a}\bar\del_{\bar a}\phi\,\Theta_-\right)
+e^{2A}\left(\bar\Theta_+\calf\del\phi\,\Theta_- +\bar\Theta_-\calf{\bar\del}\phi\,\Theta_+\right)\Big]
\end{align}
Substituting \eqref{theta plus minus} and after some more manipulations, we obtain the explicit contributions:
\begin{align}\label{S F1}
S_{\tau} &= \int\d^4\sigma\ \left(\sqrt h-\tfrac{h}{2\sqrt{M}}\right)\left[\lambda\psi^a\del_a\phi +\psi^a\rho_{ab}\bar\del^b\phi-\tilde\lambda\tilde\psi_a\bar\del^a\phi -\tilde\psi_a\tilde\rho^{ab}\del_b\phi \right]\notag\\
&\quad+\frac12\int\d^4\sigma\ e^{2A-\phi}\left[\left(\tfrac{h^{3/2}}{M}-\tfrac{h}{\sqrt{M}}\right)\psi_{\bar a}(\calf\psi)_{\bar b}\Omega^{\bar{ab{\bf v}}}\bar\del_{\bar{\bf v}}\phi
-\tfrac{h}{\sqrt M}\tilde\psi_a(\calf\tilde\psi)_b\bar\Omega^{ab{\bf v}}\del_{\bf v}\phi\right]\notag\\
&\quad-\int\d^4\sigma\  \tfrac{h}{\sqrt M}e^{4A-\phi}\left[\tilde\lambda(\calf^2\tilde\psi)_a\bar\del^a\phi-(\calf^2\psi)^a\rho_{ab}\bar\del^b\phi\right]\notag\\
&\quad-\frac12\int\d^4\sigma\ \tfrac{h^{3/2}}{M}e^{6A-2\phi}\, (\calf^2\psi)_{\bar a}(\calf \psi)_{\bar b}\Omega^{\bar{ab{\bf v}}}\bar\del_{\bar{\bf v}}\phi
\end{align}
We stress again that combining these terms with \eqref{int by part terms} and using \eqref{h and F identity}, the final expressions simplify greatly.

\subsection{Final result}
Putting together (\ref{int by part terms}, \ref{S G3 left}, \ref{S G3 right}, \ref{S F5}, \ref{S F1}) and using \eqref{h and F identity}, we finally arrive to a full expression, which we write here explicitly.
The complete action is $S^{\rm ferm}_{\rm tot}=S_{\rm kin.}+S_{\rm flux}+S'_{\rm flux}$:
\begin{align}
\label{complete final result}
S_{\rm kin.} =\ &2\int \d^4\sigma\ \sqrt h \left(\psi^a \del^+_a\lambda+\tilde\psi_a\bar\del^{-\,a}_Q\tilde\lambda-\psi^b\bar\del^{-\,a}\rho_{ab}-\tilde\psi_b\del^+_{Q\,b}\tilde\rho^{ab}\right)
\\[1em]
S_{\rm flux} =\ 
	&
	-\frac18\int \d^4\sigma\ \sqrt h\ e^{2A}\left(\rho_{ab}\rho_{cd} e^{-\phi}\calk_m{}^{e[a}\calf^{b]}{}_e \,\bar\Omega^{mcd}
	+\tilde\rho^{ab}\tilde\rho^{cd}e^{-\phi} \calk^m{}_{e[a}\calf_{b]}{}^e \,\Omega_{mcd}\right)\cr
	&+\ii\int\d^4\sigma\ \sqrt h\ e^{2A}\left(\frac14\tilde\psi_a\tilde\psi_b \bar\Omega^{ac{\bf v}} (G_{\it 2,1})_{c{\bf v}}{}^b
	- e^{2A}\tilde\psi_a \tilde\rho^{ab} (\calf\cdot G_{\it 2,1})_b\right)\notag\\
	&-\frac12\int\d^4\sigma\ \sqrt{h}\ e^{2A-\phi}\ \bar\Omega^{ab{\bf v}}\ \tilde\psi_a (\calf\tilde\psi)_b \del_{\bf v}\phi\\\notag
	&+\frac\ii8\int\d^4\sigma
	\ \sqrt h\ e^{2A}\left(\lambda\lambda(\bar{G_{\it 0,3}}\cdot\bar\Omega)
	+\frac14\rho_{ab}\rho_{cd} \bar\Omega^{ab{\bf v}}(\bar{ G_{\it 2,1}})^{cd}{}_{\bf v}\right)\cr\notag
	&-\frac\ii{16}\int \d^4\sigma\ \sqrt h\ e^{6A-\phi} 
	\left(\tilde\lambda\tilde\lambda(G_{\it 0,3}\cdot\Omega)
	+\frac14\tilde\rho^{ab}\tilde\rho^{cd} \Omega_{ab{\bf v}} (G_{\it 2,1})_{cd}{}^{\bf v}\right)\left(\calf\cdot\calf\right)
\\[1em]
\label{noncov}
S'_{\rm flux}=\ &
	\frac12\int\d^4\sigma\ \tfrac{h}{\sqrt M}e^{6A-\phi} 
	\Big(\,\d e^{-4A} -\frac\ii2\calf\cdot G_{\it 2,1} -\frac12{e^{-\phi}(\calf\cdot\calf)\del\phi} \Big)_{\bf v} \ \bar\Omega{}^{{\bf v}b}{}_{\bar a}\ \tilde\psi^{\bar a} (\calf\tilde\psi)_b \cr
	&-\int\d^4\sigma \tfrac{h}{\sqrt M}\ e^{4A-\phi}\ 
	\Big((\nabla\cdot\calf)^a-\frac\ii2 e^{\phi}\bar G{}^b{}_b{}^a\Big)\left(\tilde\lambda(\calf\tilde\psi)_a+\rho_{ab}(\calf\psi)^b \right)\\\notag
	&-\int\d^4\sigma \tfrac{h}{\sqrt M}\ e^{4A-\phi}\ \Big((\nabla\cdot\calf)^{\bar a} 
	-\frac\ii2 e^\phi G^c{}_c{}^{\bar a}\Big)\left(\lambda (\calf\psi)_{\bar a}+\tilde\rho_{\bar{ab}}(\calf\tilde\psi)^{\bar b}\right)
\end{align}

The last two rows of $S'_{\rm flux}$ vanish by imposing the standard Bianchi identity $\d \calf = H_{\it 3}|_D$. Notice that if we only imposed the equations \eqref{covBI}, which imply just a part of $\d \calf = H_{\it 3}|_D$, a term 
\begin{equation}\label{BI surviving term}
\frac14\int\d^4\sigma\, \frac h{\sqrt M}\ e^{4A}\ \ii G^c{}_c{}^{\bar a}\left(\lambda (\calf\psi)_{\bar a}+\tilde\rho_{\bar{ab}}(\calf\tilde\psi)^{\bar b}\right)
\end{equation}
would survive. However, such a contribution would generate a coupling to the universal fermionic zero-mode $\lambda_{\rm z.m.}$, even if it only involves supersymmetry-preserving fluxes. Hence, it seems justified to set these terms to zero in the derivation of the fermionic action, and relax the Bianchi identities to the form \eqref{covBI} only afterwards.%
\footnote{We could further justify this procedure noting that we are free to add terms proportional to the (standard) Bianchi identities to the Minkowskian D3-brane effective action, in such a way that they would cancel the contribution above \eqref{BI surviving term} after Wick rotation.}

All other contributions in \eqref{complete final result} can be reorganized in terms of the operators $\cals$, $\tilde\cals$, $\tilde\calu$, $\tilde\calr$ in equations (\ref{S1}, \ref{tilded operators}, \ref{calrprime}), and of $\Delta^{\rm SB}S^{\rm ferm}$ defined in \eqref{SBaction}, hence giving (\ref{fullaction}) as final result.


\section{Equations of motion}
\label{app:eom}

In general, the equations of motion are obtained by extremizing the action (\ref{E3action1}), which for the configurations we are talking about can be rewritten as
\be\label{bosaction}
S_{\rm E3}=2\pi\int_{D}\d^4\sigma\sqrt{\det(e^{-2A}h+e^{-\phi/2}\calf)}-2\pi\ii\int_{D}C\wedge e^\calf
\ee
The general equations of motion can be derived straightforwardly, see for instance \cite{Bandos:2006wb}.
However, in our case we can follow a simpler route to find them, since we are interested in the variation of $S_{\rm E3}$ around a supersymmetric configuration, which is `calibrated'
in the generalized sense of \cite{luca1}. By using the properties of generalized calibrated branes, one can then show \cite{luca2} that around a calibrated configuration the action (\ref{bosaction})  is well approximated by the action (\ref{E3action2}). More precisely, we can write
\be\label{bosaction2}
S_{\rm E3}=2\pi\int_{D}(-\frac12\, e^{-4A}J\wedge J-\frac\ii2\,\tau \calf\wedge \calf-\ii C_{\it 2}\wedge \calf-\ii C_{\it 4})+\ldots
\ee
where the missing terms are quadratic in the fields describing the deviation of the E3 configuration from the supersymmetric one.  
Hence, in order to evaluate the equations of motion on the supersymmetric configuration one can use the action (\ref{bosaction2}),
simplifying considerably the task.

A general fluctuation of the embedding is described by a section $\delta v$ of the normal bundle $N_D$ (defined by the orthogonal split $T_X|_D=T_D\oplus N_D$).  Since the embedding is holomorphic, we can split $\delta v=\delta\varphi\,{\bf v}+\delta\bar\varphi\,\bar{\bf v}$, where $\delta\varphi$ and $\delta\bar\varphi$ are complex scalars and ${\bf v}$ is a given section of $N^{1,0}_D$.  
By taking into account that $\iota_v J|_D=0$ and that $\delta_v\calf=\iota_vH|_D$, the variation of (\ref{bosaction2}), evaluated on the supersymmetry configuration, gives
\be\label{variation}
\delta_vS_{\rm E3}=2\pi\int_{D}\Big[-\iota_v\del e^{-4A}\,J\wedge J-\frac\ii2\iota_v\del\tau\, \calf\wedge\calf-\ii\,\iota_v G_{\it 3}\wedge\calf\Big]
\ee
where we have used the formula $F^{\rm int}_{2,3}=\frac{\ii}{2}\delbar e^{-4A}\, J\wedge J$, which is an alternative way of writing $F^{\rm int}_5=*_X\d e^{-4A}$, which is the internal component of (\ref{F5}). By using the fact that $*_D\calf=\calf$ and that $-\frac12\, J\wedge J=\d^4\sigma\sqrt{\det h}$, we can write (\ref{variation}) as
\be\label{variation2}
\delta_vS_{\rm E3}=2\pi\int_{D}\d^4\sigma\sqrt{\det h}\,\delta\varphi\,\Big[-8e^{-4A}\iota_{\bf v}\del A-\frac\ii2\iota_{\bf v}\del\tau\, \calf\lrcorner \calf-\ii\,\calf\lrcorner \iota_{\bf v} G_3\Big]
\ee
We see that only the component of ${\bf v}$ along the holomorphic transversal direction appears the equation of motion which is not automatically satisfied can be written as
\be\label{Xeom}
2{\bf v}(e^{-4A})-{\bf v}(\phi)e^{-\phi}\, \calf\lrcorner \calf-\ii\,\calf\lrcorner \iota_{\bf v} G_{\it 3}=0
\ee

Let us now consider a fluctuation $\delta A$ of the gauge field. Then $\delta\calf=\d\delta A$ and the variation of (\ref{bosaction2}) gives
\be\label{Avar}
\delta S_{\rm E3}=-2\pi\ii\int_{D}\delta A\wedge (G_{\it 3}+\del\tau\wedge \calf)
\ee
Now only the $({0,1})$ component of $\delta A$ appears and the non-automatically satisfied equation of motion reads
\be\label{Aeom}
G_{\it 3}|_D+\del\tau|_D\wedge \calf=0
\ee

\end{appendix}


%


\providecommand{\href}[2]{#2}\begingroup\raggedright\endgroup

\end{document}